\documentclass[prd,aps,preprintnumbers,amsmath,amssymb,nofootinbib,superscriptaddress,notitlepage]{revtex4}
\usepackage{graphicx}
\usepackage{epsfig}
\usepackage[utf8]{inputenc}
\usepackage{color}
\usepackage{epstopdf}
\usepackage[colorlinks=true,linkcolor=blue,breaklinks=true,urlcolor=blue,citecolor=blue]{hyperref}

\newcommand{\ihep}{\affiliation{Institute of High Energy Physics, Chinese Academy of Sciences, Beijing 100049, China}}
\newcommand{\ustc}{\affiliation{University of Science and Technology of China, Hefei 230026,  China}}

\begin{document}

\title{Principles and prospects of Bose-Einstein correlation study at BESIII}

\author{Hai-Ming Hu}\email{huhm@ihep.ac.cn}
\ihep

\author{Guang-Shun Huang}\email{hgs@ustc.ac.cn}
\ustc

\date{\today}

\begin{abstract}
\rule{0ex}{3ex}
One method for determining the characteristic parameters of a hadron production source is to measure the Bose-Einstein correlation functions. In this study, we present fundamental concepts and formulas related to the Bose-Einstein correlations, focusing on the measurement principles and the Lund model from an experimental perspective. We perform Monte Carlo simulations using the Lund model generator in the 2–3 GeV energy range. Through these feasibility studies, we identify key features of the Bose-Einstein correlations that offer valuable insights for experimental measurements. Utilizing data samples collected at BESIII, we perform measurements of the Bose-Einstein correlation functions, with an expected experimental precision of a few percent for the hadron source radius and incoherence parameter.
\end{abstract}

\maketitle

\section{INTRODUCTION}

The hadron production process in electron-positron collisions involves $e^+$ and $e^-$ annihilating into a virtual photon ($\gamma^\ast$), which splits into an initial quark-antiquark pair ($q\bar{q}$). The hadronization process consists of three phases: (1) perturbative Quantum Chromodynamics (pQCD) cascade evolutions of quarks and gluons, (2) hadron source formation and preliminary hadron emission, and (3) the decay of unstable hadrons.

In the energy region of the Beijing Electron Positron Collider II (BEPCII), which ranges from 2 to 5 GeV, the preliminary hadron states include continuous states and resonant states characterized by the quantum numbers $J^{PC}=1^{--}$. Because the hadronization process cannot be predicted comprehensively and quantitatively by pQCD, phenomenological models are used to describe this phase. One of the models commonly used to accomplish this task is the semi-classical Lund string fragmentation model \cite{lundmodel,bohu}. The Lund model describes the hadron source as a string, which fragments into hadrons.

The production of vector resonant states such as $\rho$, $\omega$, $\phi$, $J/\psi$, and $\psi(3686)$, can be summarized as follows: the $\gamma^\ast$ evolves into a $q\bar{q}$ pair, which constitutes the vector meson. The meson then decays into final states, a process described by the vector meson dominance model \cite{ioffe}.

\begin{figure}[htb]
    \centering
    \includegraphics[width=0.42\textwidth,height=0.18\textwidth]{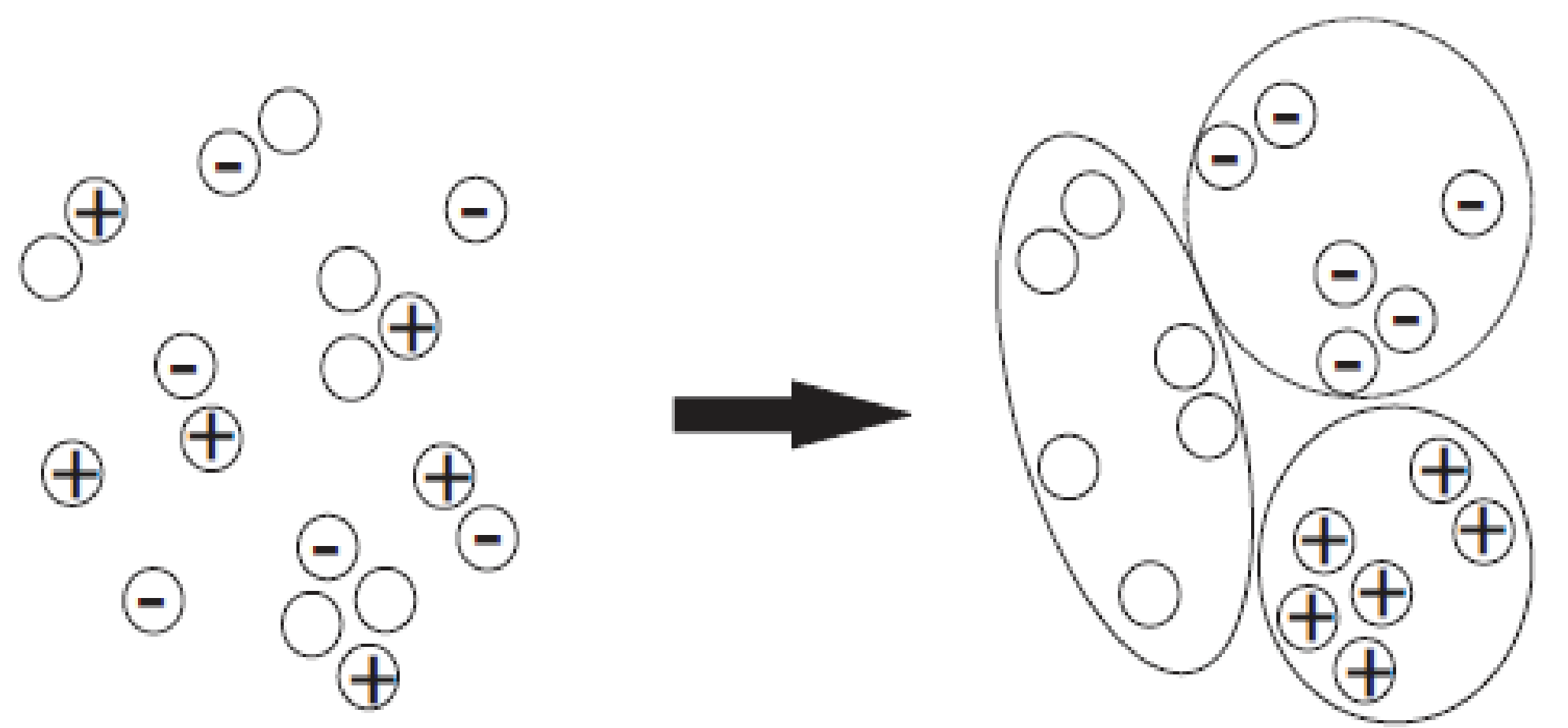}
    \caption{Identical bosons in hadron final states are more likely to be reassigned in closer phase space.}
    \label{beccondensation}
\end{figure}

In quantum mechanics, the propagation of hadrons can be described using wave functions. The wave function of a system of identical bosons is symmetric under the interchange of the space-time coordinates of any two bosons. This symmetry results in a higher probability of finding identical bosons in a small phase-space element than finding different bosons, as illustrated in Fig.\ref{beccondensation}. This phenomenon is known as Bose-Einstein correlation (BEC), which is independent of specific interactions. Measurements of the BEC effect provide valuable space-time information about the hadron source \cite{boal,weiner}.
\vspace{-2mm}

Theoretical studies of the BEC effect have been conducted for many years, leading to various models of the space-time distribution of the hadron source, including Gaussian- and ellipsoid-type distributions \cite{boal,weiner}, as well as the more generalized  L$\acute{e}$vy distribution \cite{phenix,adare,csorgo}.

Numerous collaborations have conducted experimental measurements of the BEC effect involving $e^+e^-$, $p\bar{p}$ collisions, and $ep$ and $Kp$ scattering to select various signal samples at energies ranging from tens of GeV to TeV. A selection of these studies are listed in Table 1, and additional details on other experiments are documented in Ref.\cite{boutemeur}. All these experiments demonstrate the universal characteristics of the BEC effect.
\begin{center}
\begin{tabular}{lccc}\hline\hline
Collaboration           & Beam     &$\sqrt{s}$ {\small(GeV)} & Boson pairs                                 \\\hline
MARKII \cite{markii}    & $e^+e^-$        &29               & $\pi^\pm\pi^\pm$                            \\
AMY    \cite{amy}       & $e^+e^-$        &58               & $\pi^\pm\pi^\pm$                            \\
OPAL   \cite{opal}      & $e^+e^-$        &91               & $\pi^\pm\pi^\pm$, $\pi^0\pi^0$              \\
L3     \cite{l3,achard} & $e^+e^-$        &91               & $\pi^0\pi^0$                                \\
NA22   \cite{na22}      & $\pi^+p$,$K^+p$ &21.7             & $\pi^\pm\pi^\pm$                            \\
ZEUS   \cite{epzeus}    & ep              &10.5             & $\pi^\pm\pi^\pm$,$K^\pm K^\pm$              \\
CMS    \cite{cms}       & pp              &900,2360         & $\pi^\pm\pi^\pm$                            \\
ATLAS  \cite{atlas}     & pp              &13000            & $\pi^\pm\pi^\pm$                            \\
ALICE  \cite{alice}     & pp              &900,7000         & $\pi^+\pi^+$                                \\
PHENIX \cite{phenix}    & AuAu            &200              & $\pi^\pm\pi^\pm$                            \\
\hline\hline
\end{tabular}\\
\vspace{0.75mm}
Table 1. Status of measurements of BEC effect.
\end{center}

However, in the BEPCII energy region, measurements of the BEC effect are still lacking. The BESIII detector at BEPCII has collected data samples with large-sample statistics at certain energies, providing the opportunity to measure the BEC effect. In this study, we first introduce the basic concepts of the BEC effect. Then we derive the emission amplitude of the hadron source based on the Lund model, as well as the wave functions of a two-pion system. Finally, we perform feasibility studies using the Monte Carlo (MC) method via the Lund model generator to explore the properties relevant for measuring the BEC effect.

\section{HADRON SOURCE IN THE LUND MODEL}

\subsection{Hadron production at space-time pointS}

\begin{figure}[htb]
    \centering
    \includegraphics[width=0.4\textwidth]{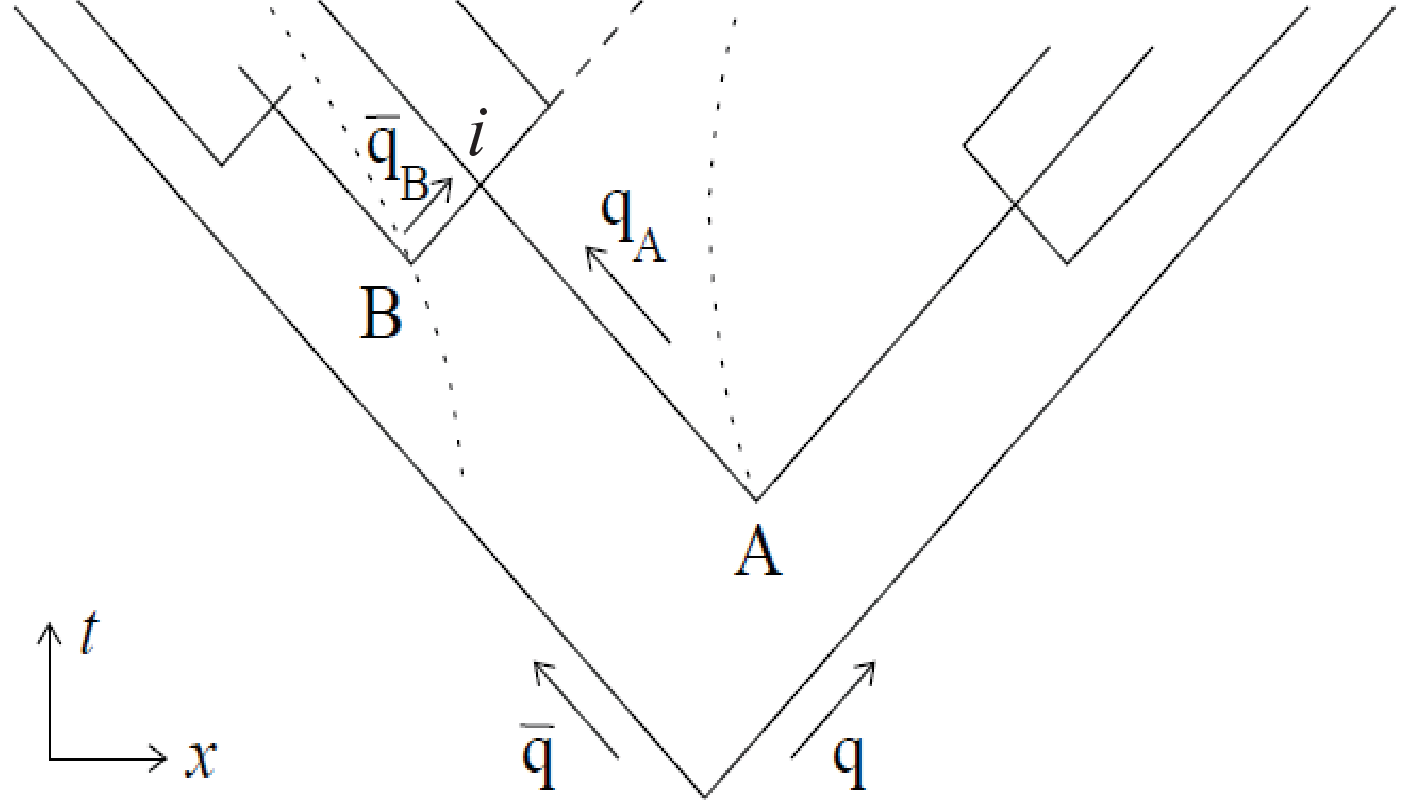}
    \caption{Meson $i$ with constituent quarks $q_A$ and $\bar{q}_B$ lying on hyperbolas of space-time.}
    \label{hyperbolas}
\end{figure}

The hadron source in the Lund model is not static, and it evolves in space-time according to the Lund area law \cite{bohu}. Hadrons are emitted from convex vertices at various space-time points via string fragmentation. The mass and momentum of a hadron originate from the potential energy stored in the string. Figure \ref{hyperbolas} shows that meson $i$ is produced at space-time point $i=(x_i,t_i)$, and is composed of $q_A$ and $\bar{q}_B$ located at adjacent concave vertices A=$(x_A,t_A)$ and B=$(x_B,t_B)$. The mass and energy-momentum of meson $i$ satisfy the relativistic mass-energy relation, which is expressed as \cite{lundmodel}
\begin{equation}\label{meprelation}
\frac{E_i^2-p_i^2}{\kappa^2}=\frac{m_i^2}{\kappa^2}=(x_A-x_B)^2-(t_A-t_B)^2,
\end{equation}
where $\kappa$ is the attractive tension constant between quarks and antiquarks.

Equation (\ref{meprelation}) indicates that points A and B lie on two hyperbolas, and that the space-time interval between A and B should be large enough to ensure that the produced hadron has the required mass. Consequently, the hadron source in the Lund model is discontinuous in space-time.

The Lund area law provides a strict solution for the Lund model. The probability of producing $n$ hadrons is given by \cite{lundmodel,bohu}
\begin{equation}\label{dpn}
d{\cal P}_n(\{p_j\})=\delta^2(P_n-\sum_{j=1}^np_j)\prod_{j=1}^nd^2p_j\delta(p_j^2-m_j^2)|{\cal M}_n|^2,
\end{equation}
where $p_j$ is the momentum component of the hadron parallel to the string, and $m_j$ is the transverse mass. The matrix element ${\cal M}_n$ is expressed as
\begin{equation}
{\cal M}_n=\exp[i\xi{\cal A}_n],~~\xi=1/2\kappa+ib/2
\end{equation}
where ${\cal A}_n$ is the light-cone area, and $b$ is a dynamics constant.

The space-time $\{x_i\}$ and energy-momentum $\{p_i\}$ are related through Eq.(\ref{meprelation}), implying that these sets of quantities are equivalent. Thus, ${\cal M}_n$ can be viewed as a function of either $\{x_i\}$ or $\{p_i\}$ equivalently, which can be expressed as
\begin{equation}
{\cal M}_n={\cal M}_n(\{p_i\};\{x_i\})~~{\rm or}~~{\cal M}_n={\cal M}_n(\{x_i\};\{p_i\}),
\end{equation}
where the quantities in the first set of braces are the variables of ${\cal M}_n$, and those in the second set of braces are the conjugate parameters. In principle, using the correspondence between $\{x_i\}$ and $\{p_i\}$ and the Jacobian transformation, $d{\cal P}_n(\{p_j\})$ in Eq.(\ref{dpn}) can be rewritten as the conjugate expression $d{\cal P}_n(\{x_j\})$ in space-time coordinates, which is the space-time distribution of the hadron emission source.

\subsection{Emitting amplitude of hadron source}

To match experimental data, the Lund model incorporates phenomenological parameters that represent hadron production ratios at each vertex \cite{jetset,bohu}. Consequently, each matrix element ${\cal M}_n$ is associated with a set of additional parameters that depend on the hadron type, which is denoted as the factor $g_n$. The exclusive hadron emitting amplitude $F_n(\{x_i\};\{p_i\})$ can be expressed as
\begin{equation}\label{Fnxipi}
F_n(\{x_i\};\{p_i\})\sim g_n{\cal M}_n(\{x_i\};\{p_i\}).
\end{equation}

The emitting amplitude of a single particle produced at $x_i$ with momentum $p_i$ in an event of multiplicity $n$ can be expressed as
\begin{equation}\label{fnxipi}
f_n(x_i;p_i)\sim\sum_{j\not=i}F_n(\{x_j\};\{p_j\}),
\end{equation}
which is a complex number and depends on the hadron category.

The quantity $f_n(x_i;p_i)$ shown in Eq.(\ref{fnxipi}) is impossible to calculate analytically. The single-particle emitting probability amplitude is numerically obtained from MC samples with a fixed $n$ via
\begin{equation}\label{fnxipimc}
f_n(x_i;p_i)\sim\frac{1}{N_{evt}}\sum_{ievt}^{Nevt}\sum_{j\not=i}F_n(\{x_j\};\{p_j\}),
\end{equation}
where $N_{evt}$ is the number of events in the MC sample. The inclusive single-hadron emitting probability amplitude in MC simulations corresponds to
\begin{equation}\label{fxipimc}
f(x_i;p_i)\sim\frac{1}{N_{evt}}\sum_{ievt}^{Nevt}\sum_{n}\sum_{j\not=i}F_n(\{x_j\};\{p_j\}).
\end{equation}

\section{WAVE FUNCTION OF A TWO-PION SYSTEM}

In relativistic terms, a free and stable particle can be described by a plane wave function (strictly speaking, however, the motion of particles should be described by wave packets):
\begin{equation}
\psi(x;p)\sim\exp(ipx).
\end{equation}
The lifetime of certain hadrons, such as $\pi^\pm$ and $K^\pm$, are sufficiently long to be considered stable in experiments. For an unstable particle with a short life-time (resonance), the wave function can be written as
\begin{equation}\label{rlbwwf}
\psi(x;p)\sim\frac{m\Gamma}{s-m^2+im\Gamma}\exp(ipx),
\end{equation}
where $m$ is the resonance peak position (the mass of the resonance), $s$ is the square of the center-of-mass energy, and $\Gamma$ is the decay width (which indicates the lifetime of the particle via $\tau=\hbar/\Gamma$. For simplicity, we neglect the phase factor $\exp(i\theta)$ as it does not affect the analysis of the BEC effect.

However, if a particle is influenced by Coulomb and strong interactions, the corresponding wave function becomes significantly more complex \cite{hepph9606365}. This study does not discuss the situation of final state interaction.

For the BESIII detector, the spatial coordinate origin $O$ was defined as the center of the detector, which also served as the collision point of $e^+e^-$, and the time origin was synchronized with the event trigger clock of the data acquisition system.

\begin{figure}[htb]
    \centering
    \includegraphics[width=0.4\textwidth,height=0.25\textwidth]{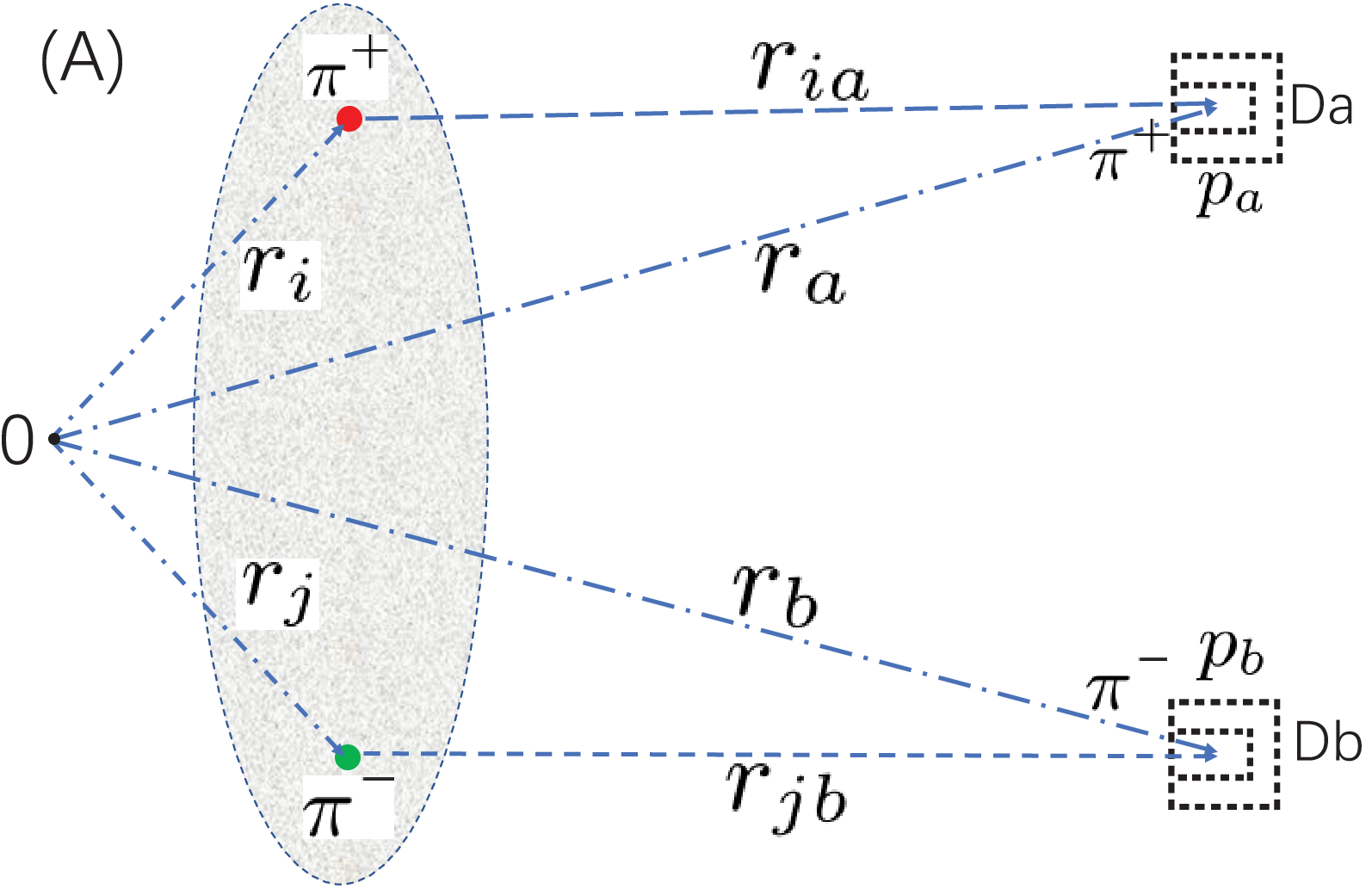}
    \includegraphics[width=0.4\textwidth,height=0.25\textwidth]{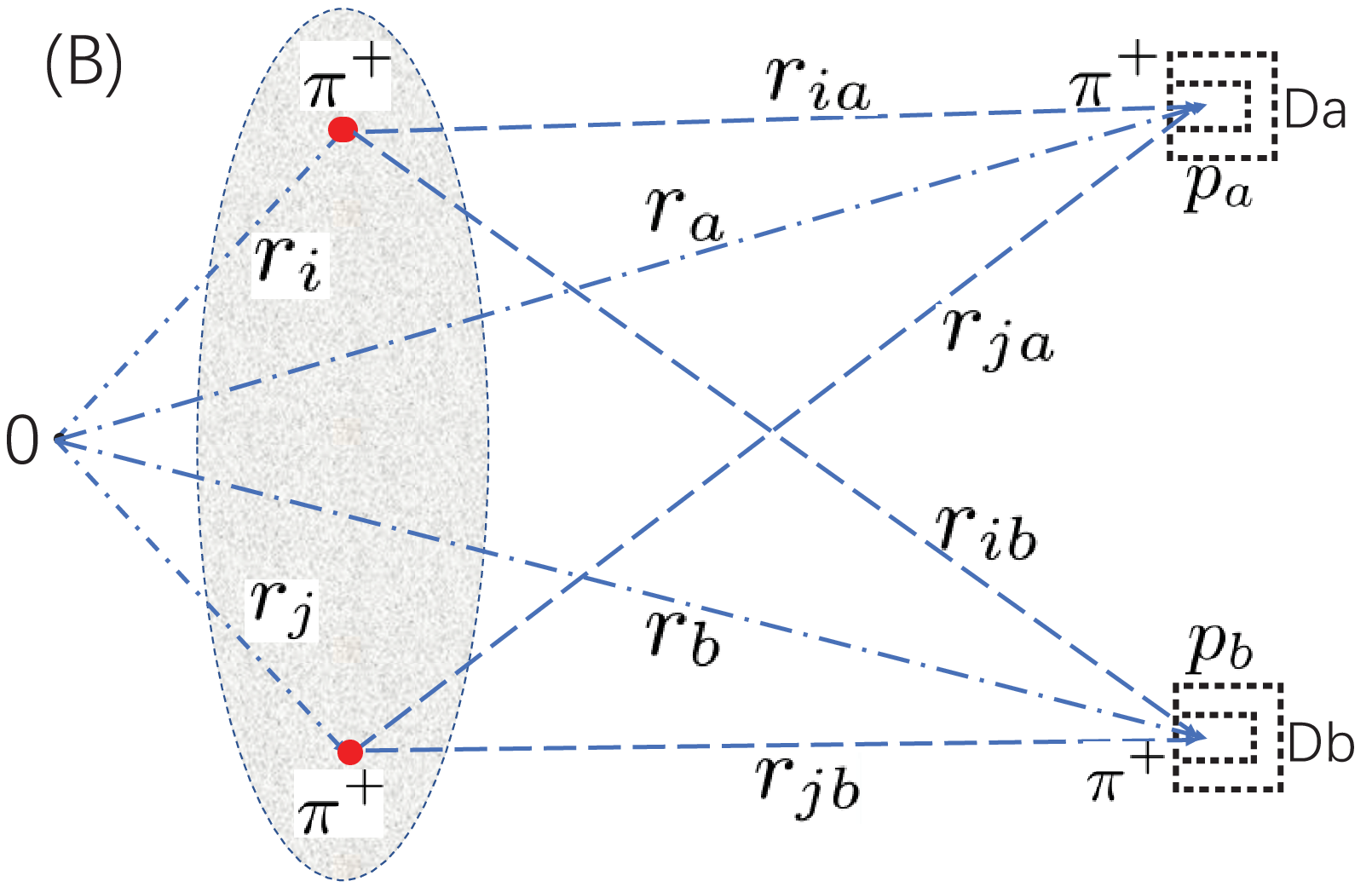}
    \caption{(color online) $\pi^+\pi^-$ and $\pi^+\pi^+$ are emitted from the hadron source and detected by detectors Da and Db.}
    \label{hadronscouce}
\end{figure}

As shown in Fig.\ref{hadronscouce}, two pions ($\pi^+\pi^-$ and $\pi^\pm\pi^\pm$) are emitted from the hadron source points at $x_i=({\bf r_i},t_i)$ and $x_j=({\bf r_j},t_j)$, respectively, and are detected by detectors Da at $x_a=({\bf r_a},t_a)$ and detector Db at $x_b=({\bf r_b},t_b)$. Here, ${\it\bf r}_a$ and ${\it\bf r}_b$ are the central positions of Da and Db relative to $O$, respectively, and $t_a$ and $t_b$ represent the times required for particles to travel from $O$ to Da and Db, respectively.

For a real detector, Da is not a point but has a finite range, and time and momentum are measured with uncertainties. Thus, the detected interval $x_{ia}$ and $p_a$ can be expressed as
\begin{equation}
x_{ia}=x_a-x_i+\delta x_{ia}~~{\rm and}~~p_a=p_i+\delta p_a,
\end{equation}
where $\delta x_{ia}$ and $\delta p_a$ denote the detected space-time and energy-momentum uncertainties. The wave function of $\pi^+$ arriving at $x_a$ is given by
\begin{equation}\label{psiia}
\psi(i;a)\equiv\psi(x_{ia};p_a)=\exp(ip_ax_{ia})=\exp(ip_ax_a)\times\exp(-ip_ax_i+i\delta_{ia}),
\end{equation}
where $\delta_{ia}=p_a\delta x_{ia}+\delta p_ax_{ia}$ is a random phase angle, possibly with observable effects on the measurement of the correlation functions.

Similarly, the wave function of $\pi^-$ emitted from $x_j$ with momentum $p_j$ and detected by Db at $x_b$ with momentum $p_b$ is given by
\begin{equation}\label{psijb}
\psi(j;b)\equiv\psi(x_{jb};p_b)=\exp(ip_bx_{jb})=\exp(ip_bx_b)\times\exp(-ip_bx_j+i\delta_{jb}),
\end{equation}
where $x_{jb}=x_b-x_j+\delta x_{jb}$, $p_b=p_j+\delta p_b$, and $\delta_{jb}=p_b\delta x_{jb}+\delta p_bx_{jb}$. The joint wave function of $\pi^+\pi^-$ arriving at Da and Db is expressed as
\begin{equation}
\psi_d(i,j;a,b)\equiv\psi(i,a)\psi(j,b)=\psi(x_{ia};p_a)\psi(x_{jb};p_b).
\end{equation}
This expression for $\psi_d$ also applicable to $\pi^{\pm}\pi^0$.

As depicted in Fig.\ref{hadronscouce}(B), $\pi^+$ and $\pi^+$ are detected by Da and Db with momenta $p_a$ and $p_b$ respectively. The two identical pions are indistinguishable according to measurements, meaning Da and Db cannot determine whether a $\pi^+$ was emitted from $x_i$ or $x_j$. Consequently, the joint wave function of $\pi^+\pi^+$ has a symmetrized form for interchanging source points
$i$ and $j$:
\begin{equation}\label{psis}
\psi_s(i,j;a,b)\equiv\frac{1}{\sqrt{2}}[\psi(i;a)\psi(j;b)+\psi(j;a)\psi(i;b)].
\end{equation}
The expression for $\psi_s$ is also applicable to $\pi^-\pi^-$ and $\pi^0\pi^0$.

If the preliminary state contains unstable particles that decay into stable final states, owing to the fact that the lifetimes of most unstable particles are only a few femtometer (much smaller than the spatial resolution of the detector), the decay vertex is considered as part of the experimental uncertainty in the emitting position.

\section{DETECTION PROBABILITY AMPLITUDE}

The inclusive detection probability amplitudes for $\pi^+$ and $\pi^-$, emitted from points $x_i$ and $x_j$ and detected by detectors Da and Db with momenta $p_a$ and $p_b$, respectively, are expressed as
\begin{equation}\label{aria}
{\cal A}(i;a)\equiv{\cal A}(x_{ia};p_a)=f(x_i;p_i)\psi(x_{ia};p_a),
\end{equation}
\begin{equation}\label{arjb}
{\cal A}(j;b)\equiv{\cal A}(x_{jb};p_b)=f(x_j;p_j)\psi(x_{jb};p_b).
\end{equation}
In these calculations, we assume that $p_a=p_i$ and $p_b=p_j$ (incorporating measurement uncertainties into the random phase angles $\delta_{ia}$ and $\delta_{jb}$), and that they can be absorbed into the functions $f(x_i;p_a)$ and $f(x_j;p_b)$, respectively. Therefore, the detection probability amplitudes not only encode information about the hadron sources but are also influenced by measurement uncertainties.

The joint probability amplitude that $\pi^+$ and $\pi^-$ will be detected by Da and Db is given by
\begin{equation}\label{obsampd}
{\cal A}_d(i,j;a,b)\equiv {\cal A}(i;a){\cal A}(j;b).
\end{equation}

For the case in which two identical $\pi^+\pi^+$ are detected by Da and Db, four probability amplitudes exist, as shown in Fig.\ref{hadronscouce}(B). Besides the two already given by Eqs.(\ref{aria}) and (\ref{arjb}), the additional amplitudes are
\begin{equation}\label{arja}
~~{\cal A}(j,a)\equiv{\cal A}(x_{ja};p_a)=f(x_j;p_j)\psi(x_{ja};p_a),
\end{equation}
\begin{equation}\label{arib}
{\cal A}(i,b)\equiv{\cal A}(x_{ib};p_b)=f(x_i;p_i)\psi(x_{ib};p_b),
\end{equation}
The joint probability amplitude for detecting $\pi^+\pi^+$ is then
\begin{equation}\label{obsamps}
{\cal A}_s(i,j;a,b)\equiv\frac{1}{\sqrt{2}}[{\cal A}(i,a){\cal A}(j,b)+{\cal A}(j,a){\cal A}(i,b)],
\end{equation}
which is symmetric with respect to the interchange of source points $i$ and $j$.

\section{CORRELATION FUNCTION OF TWO PIONS}

The expression for the correlation function for a two-pion system varies depending on whether the measurement is coherent or incoherent.

\subsection{Coherent and incoherent measurements}

One of the key considerations in BEC effect measurements is determining whether the measurement is coherent or incoherent. These two types of measurements are represented by different formulas, leading to different results. In other words, the results of the measurement are influenced not only by the detected particles but also by the measurement method and the status of the detector. A more detailed explanation of how to distinguish between coherent and incoherent measurements can be found in Ref.\cite{boal}.

The principles determining whether a measurement of the BEC effect is coherent or incoherent are analogous to those of the well-known electron double-slit experiments. If the experiment does not detect the slit through which the electron passes, the screen will display interference (coherent) patterns. Conversely, if the electron's path is detected, the screen will show non-interferential (incoherent) patterns. Similar results have been observed in double-slit experiments using neutrons and atoms, indicating the universal quantum characteristics of microscopic particles.

\begin{figure}[htb]
    \centering
    \includegraphics[width=0.325\textwidth]{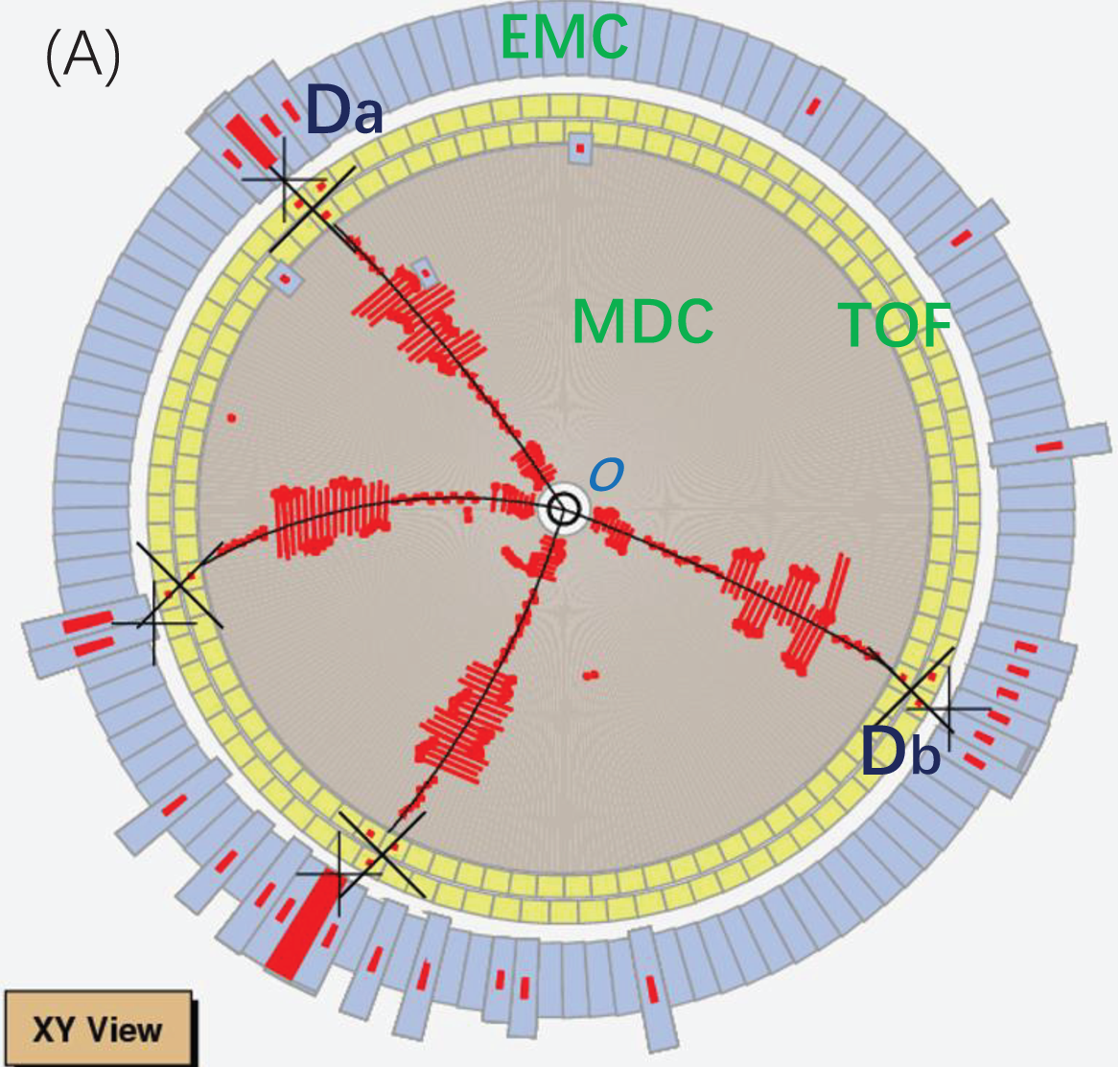}
    \includegraphics[width=0.325\textwidth]{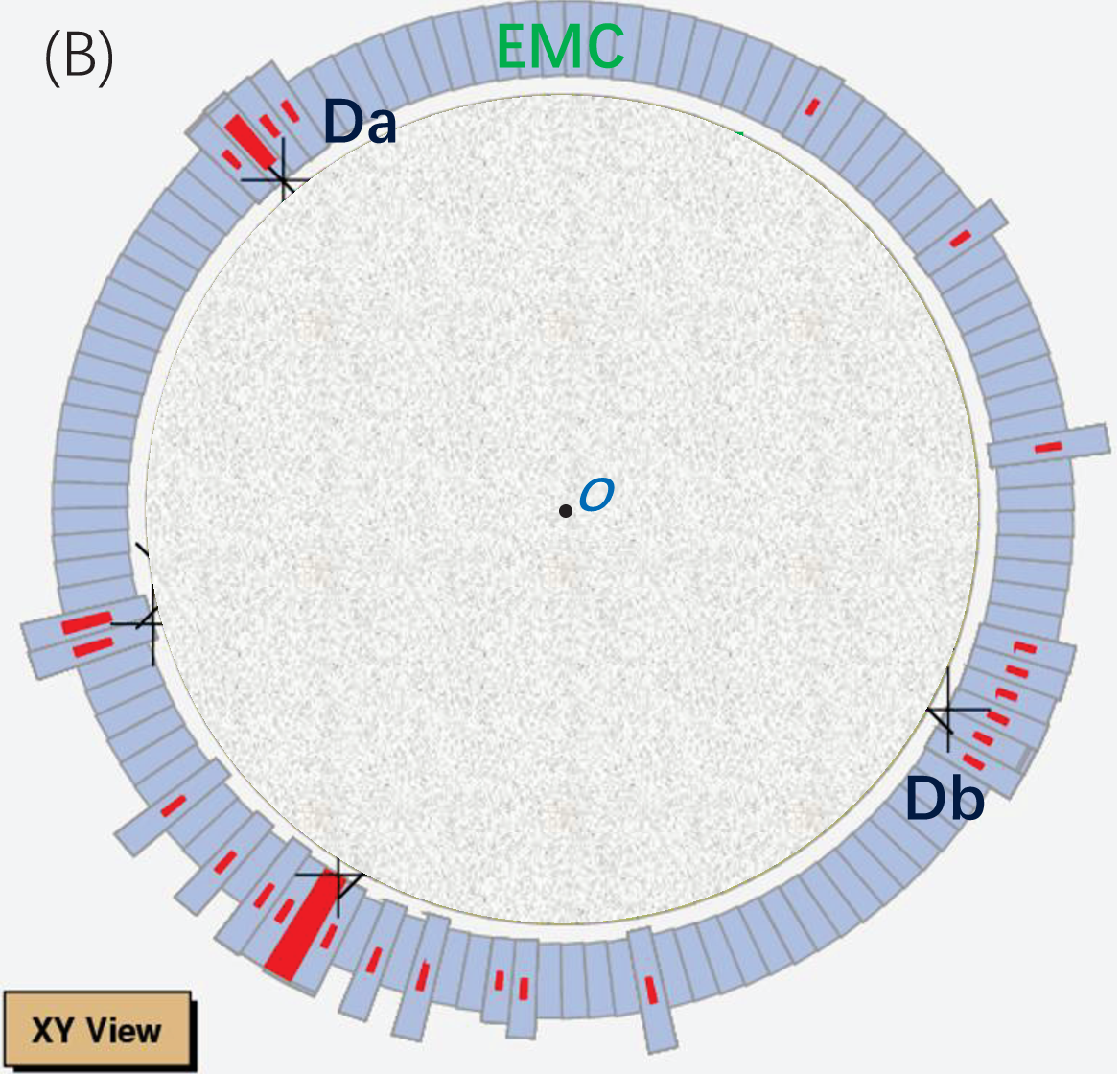}
    \caption{(color online) Illustrations of two kinds of detectors. Detector (A) can determine from where the particles were emitted, but detector (B) can-not.}
    \label{BESIII}
\end{figure}

Based on the principles discovered in double-slit experiments, we can define the criteria for a BEC measurement to be coherent or incoherent as follows:

$\bullet$ Incoherent measurement: If the detector can determine the emission position and the trajectory of a particle, such as those found by the BESIII detector shown in Fig.\ref{BESIII}(A), using a vertex detector (VD) or track reconstruction in the main drift chamber (MDC), the measurement is considered incoherent. In this scenario, the detection of particle paths eliminates interference effects.

$\bullet$ Coherent measurement: If the detector cannot determine the emission position and trajectory, as reflected by the setup shown in Fig.\ref{BESIII}(B), and the VD and MDC are turned off and only the electromagnetic calorimeter (EMC) is operational, the measurement is considered coherent. In this case, the inability to track the particle paths allows interference patterns to emerge.

For a real detector that falls somewhere between the status of Fig.\ref{BESIII}(A) and Fig.\ref{BESIII}(B), the detected correlation functions should be a combination of both incoherent and coherent terms. This mixed measurement results from partial information about particle trajectories, leading to partial interference effects.

\subsection{Incoherent correlation distributions}

When measurements are performed using data collected with an ideal and perfect detector, which is shown in Fig.\ref{BESIII}(A), the measurement is guaranteed to be incoherent. This incoherence is a consequence of the inherent quantum characteristics of the system.

$\bullet$ Distribution of a single pion: When a hadron source emits a pion from position $x_i$ with momentum $p_i$, and it is detected by detector Da with momentum $p_a$, the inclusive probability distribution is given by
\begin{equation}\label{p1i}
P_1^I(p_a)=\sum_i|{\cal A}(i;a)|^2=\sum_i|f(x_i;p_a)|^2,
\end{equation}
which depends on the space-time distribution of only the hadron source.

$\bullet$ Distribution of two different pions: For a pair of $\pi^+\pi^-$ emitted from position $x_i$ with momentum $p_i$ and position $x_j$ with momentum $p_j$, and detected by Da with momentum $p_a$ and Db with momentum $p_b$, the incoherent correlation distribution is expressed as
\begin{equation}\label{Pdisum}
P_{2d}^I(p_a,p_b)=\sum_{i,j}|{\cal A}_d(i,j;a,b)|^2=P_1^I(p_a)P_1^I(p_b)=P_2^I(p_a,p_b).
\end{equation}
This equation shows that the incoherent correlation distribution for different pion types is the product of individual single-pion probability distributions. The same expression also applies to pairs of $\pi^\pm\pi^0$.

$\bullet$ Distribution of two identical pions: For a pair of $\pi^+\pi^+$, emitted from position $x_i$ with momentum $p_i$ and position $x_j$ with momentum $p_j$, and detected by Da with momentum $p_a$ and by Db with momentum $p_b$, the incoherent correlation distribution is
\begin{equation}\label{Psisum}
P_{2s}^I(p_a,p_b)=\frac{1}{2}\sum_{i,j}|{\cal A}_s(i,j;a,b)|^2=P_2^I(p_a,p_b)+P_s^{I}(p_a,p_b),
\end{equation}
where the additional term is defined as:
\begin{equation}\label{TsIpapb}
P_s^{I}(p_a,p_b)=\sum_{i,j}{\cal A}^\ast(i;a){\cal A}^\ast(j;b){\cal A}(i;b){\cal A}(j;a).
\end{equation}

In Eq.(\ref{Psisum}), the first term represents the genuine incoherent correlation distribution, which is similar to $P_{2d}^I(p_a,p_b)$ in Eq.(\ref{Pdisum}). The second term, $P_s^{I}(p_a,p_b)$, represents the interference, which accounts for phase averaging between the detection amplitudes of the two $\pi^+$ in  ${\cal A}_s$ given by Eq.(\ref{obsamps}). The interference term $P_s^{I}(p_a,p_b)$ in Eq.(\ref{TsIpapb}) disappears if the phases $\{\delta_{ia}\}$ and $\{\delta_{jb}\}$  in the wave functions described by Eq.(\ref{psiia}) and (\ref{psijb}) fluctuate randomly. The same expressions also apply to pairs of $\pi^-\pi^-$ and $\pi^0\pi^0$.

\subsection{Coherent correlation distributions}

When data samples are collected using the detector shown in Fig.\ref{BESIII}(B), the measurements are coherent.

$\bullet$ Distribution of a single particle: The inclusive distribution of a pion emitted from $x_i$ with  momentum $p_i$ and detected by Da with momentum $p_a$ is given by
\begin{equation}\label{p1c}
P_1^C(p_a)=|\sum_i{\cal A}(i;a)|^2=P_1^I(p_a)+\sum_{i\not=j}{\cal A}^\ast(i;a){\cal A}(j;a).
\end{equation}
In this expression, the first term $P_1^I(p_a)$, given by Eq.(\ref{p1i}), represents the incoherent contribution, which depends on the intensity of individual sources, and the second term represents the coherent contribution, which reflects the interference between different source points. In the summation of these cross terms, each amplitude has a corresponding conjugate term, and their imaginary parts cancel each other out, leaving only the real parts.

$\bullet$ Correlation between two different pions: For a pair of $\pi^+$ and $\pi^-$ detected by Da with momentum $p_a$ and by Db with momentum $p_b$, the coherent correlation distribution is
\begin{equation}\label{Pdcsum}
P_{2d}^C(p_a,p_b)=|\sum_{i,j}{\cal A}(i;a){\cal A}(j;b)|^2=P_2^I(p_a,p_b)+P_d^{C}(p_a,p_b),
\end{equation}
where
\begin{equation}
P_d^{C}(p_a,p_b)=\sum_{i\not=k,j\not=l}{\cal A}^\ast(i;a){\cal A}^\ast(j;b) {\cal A}(k;a){\cal A}(l;b).
\end{equation}
In Eq.(\ref{Pdcsum}), the first term is the incoherent correlation function given by Eq.(\ref{Pdisum}), and the second term, $P_d^{C}(p_a,p_b)$, is the true coherent one.

$\bullet$ Correlation between two identical pions: For a pair of identical $\pi^+\pi^+$ detected by Da with momentum $p_a$ and by Db with momentum $p_b$, the coherent correlation distribution is
\begin{equation}\label{Pscsum}
P_{2s}^C(p_a,p_b)=|\sum_{i,j}{\cal A}_s(i,j;a,b)|^2=P_2^I(p_a,p_b)+P_s^I(p_a,p_b)+P_s^C(p_a,p_b),
\end{equation}
where $P_s^I(p_a,p_b)$ is given by Eq.(\ref{TsIpapb}), and
\begin{equation}
P_s^C(p_a,p_b)=\sum_{i\not=k,j\not=l}{\cal A}^\ast(k;a){\cal A}(l;a){\cal A}^\ast(j;b){\cal A}(l;b)
\end{equation}
is the true coherent term of the symmetric wave function. The sign (positive or negative) of the summation of the cross terms is determined by the average of the random phases in Eqs.(\ref{psiia}) and (\ref{psijb}). This inherently quantum property produce experiments-dependent measurement results owing to the random phase angles caused by the experimental uncertainties.

Many factors contribute to correlation effects in the hadron final state. For instance, the $\delta$ function in Eq.(\ref{dpn}) imposes strong restrictions on the hadron distribution, leading to energy-momentum correlations among all final state particles. The correlation effect caused by energy-momentum conservation depends on the topology of the event multiplicity. To avoid bias, both the signal and the reference samples should be selected from events with the same multiplicity.

\subsection{Correlation function reflecting BEC effect}

We now focus exclusively on the BEC effect. The BEC function (BECF) in experiments is defined as the ratio of the correlation functions of identical bosons (the signal sample) to the correlation functions of different bosons (the reference sample). This ratio isolates the BEC effect by counteracting other correlation effects.

In an ideal scenario, if data samples of $\pi^+\pi^+$ and $\pi^+\pi^-$ are collected using a perfect detector, as depicted in Fig.\ref{BESIII}(A), this is termed an ideal measurement, because the precise information of the hadron source is detected. The ratio of the corresponding incoherent correlation functions, based on Eqs.(\ref{Pdisum}) and (\ref{Psisum}), is given by
\begin{equation}\label{c2i}
C_2^I(p_a,p_b)=\frac{P_{2s}^I(p_a,p_b)}{P_{2d}^I(p_a,p_b)}=1+R_2^I(p_a,p_b),
\end{equation}
where,
\begin{equation}\label{r2i}
R_2^I(p_a,p_b)=\frac{P_s^{I}(p_a,p_b)}{P_d^I(p_a,p_b)}.
\end{equation}

In a more realistic scenario, if data samples of $\pi^+\pi^+$ and $\pi^+\pi^-$ are collected using an imperfect detector, as shown in Fig.\ref{BESIII}(B), this is termed a non-ideal measurement because no information about the hadron emission source is detected. In this case, the measurement is coherent, and the ratio of their correlation functions, based on Eqs.(\ref{Pdcsum}) and (\ref{Pscsum}) is given by:
\begin{equation}\label{c2c}
C_2^C(p_a,p_b)=\frac{P_{2s}^C(p_a,p_b)}{P_{2d}^C(p_a,p_b)}=1+R_2^C(p_a,p_b),
\end{equation}
where
\begin{equation}\label{r2c}
R_2^C(p_a,p_b)=\frac{P_s^{C}(p_a,p_b)}{P_d^C(p_a,p_b)}.
\end{equation}

Ideally, the data collected by the detector shown in Fig.\ref{BESIII}(A) would be used to measure the BECF. However, real detectors are imperfect, often because of factors such as electromagnetic noises in the MDC or an incomplete reconstruction of the particle tracks, which indicates emitting points of the hadron source are not detected. Thus, the measured BECF is expressed as
\begin{equation}\label{realC2}
C_2(p_a,p_b)=\frac{P_{2s}(p_a,p_b)}{P_{2d}(p_a,p_b)},
\end{equation}
where $P_{2s}(p_a,p_b)$ and $P_{2d}(p_a,p_b)$ are combinations of incoherent and coherent correlation functions of the signal and reference samples respectively:
\begin{equation}\label{realP2s}
P_{2s}(p_a,p_b)=P_{2s}^I(p_a,p_b)+\alpha P_{2s}^C(p_a,p_b),
\end{equation}
\begin{equation}\label{realP2d}
P_{2d}(p_a,p_b)=P_{2d}^I(p_a,p_b)+\beta P_{2d}^C(p_a,p_b).
\end{equation}
The quantities $\alpha$ and $\beta$ reflect the degree of imperfection in the detector, which causes some of hadron emissions to be measured in a coherent-like manner. Typically,
$\alpha$ and $\beta$ are small, and the first-order approximation is given by
\begin{equation}\label{realC2approx}
C_2(p_a,p_b)=1+\lambda^{CI}R_2^I(p_a,p_b),
\end{equation}
where $R_2^I(p_a,p_b)$ is the ideal two-pion correlation function defined in Eq.(\ref{r2i}), and
\begin{equation}\label{lambdaci}
\lambda^{CI}=1+\alpha R_{2s}^{CI}(p_a,p_b)-\beta R_{2d}^{CI}(p_a,p_b),
\end{equation}
with
\begin{equation}
R_{2s}^{CI}(p_a,p_b)=\frac{P_{2s}^C(p_a,p_b)}{P_{2s}^I(p_a,p_b)}~~~~{\rm and}~~~~R_{2d}^{CI}(p_a,p_b)=\frac{P_{2d}^C(p_a,p_b)}{P_{2d}^I(p_a,p_b)},
\end{equation}
 where $\lambda^{CI}$ represents the deviation of the measured BECF from the ideal BECF owing to the imperfections of the real detector. The coefficient $\lambda^{CI}$ is distinct from the incoherent parameter $\lambda$ in phenomenological hadron source models and is adopted to fit the experimental data, which arises from different factors \cite{boal}.

Currently, two main approaches for describing the origin of the BEC effect exist: the wave function approach and the more general field theory approach. The wave function approach is intuitive for explaining the BEC effect, as it assumes that the BECF can appear coherent or incoherent depending solely on the measurement method and experimental equipment, regardless the coherence of the hadron source. In contrast, the field theory approach assumes that the hadron source is divided into incoherent and coherent field components  \cite{boal,weiner,csorgo}.

\section{BEC EFFECT IN THE LUND MODEL}

The BEC effect is a natural consequence of the Lund model \cite{lundmodel,npb5131998627}. In this model, multiple string configurations can fragment into the same hadron state. For example, as shown in Fig.14.1 of Ref.\cite{lundmodel}, there are two configurations with light-cone areas ${\cal A}_{12}$ and ${\cal A}_{21}$, both of which fragment into the same hadron final state but with different momenta. In these configurations, two identical bosons are produced at convex vertices 1 and 2. The difference between these two configurations is simply the exchange of the two identical bosons. The total matrix element for this hadron state can be expressed as
\begin{equation}\label{m12m21}
{\cal M}={\cal M}_{12}+{\cal M}_{21}
\end{equation}
leading to
\begin{equation}
|{\cal M}|^2=[\exp(-b{\cal A}_{12}+\exp(-b{\cal A}_{21})](1+{\cal H}),
\end{equation}
where the term ${\cal H}$ is defined as
\begin{equation}
{\cal H}=\frac{\cos(\Delta{\cal A}/\kappa)}{\cosh(b\Delta {\cal A}/2)},
\end{equation}
where $\Delta{\cal A}$ is the difference between ${\cal A}_{12}$ and ${\cal A}_{21}$. This shows that the interference between the two configurations ${\cal A}_{12}$ and ${\cal A}_{21}$ results in an enhancement factor of $1+{\cal H}$. This factor increases as $\Delta{\cal A}$ decreases, which occurs when the momenta $p_1$ and $p_2$ of the two identical bosons become more similar in magnitude.

\section{BECF IN EXPERIMENTAL MEASUREMENTS}

In quantum field theory, the distribution functions are related to the differential cross sections. For the inclusive process $a+b\to c+X$ (where $X$ represents all other particles except $c$), the differential cross section is defined as
\begin{equation}
d\sigma(p)=\frac{\sigma_{tot}}{<n>}\tilde{\rho}_1(p)d\phi=\sigma_{tot}\rho_1(p)d\phi,
\end{equation}
where $d\phi$ is the phase-space element of particle $c$, $<n>$ is the average multiplicity, and $\rho_1(p)$ is the single-particle density distribution. Similarly, for the two-particle differential cross section in the process $a+b\to c_1+d_2+X$, we have
\begin{equation}
d^2\sigma(p_1,p_1)=\sigma_{tot}\rho_2(p_1,p_2)d\phi_1d\phi_2,
\end{equation}
where $\rho_2(p_1,p_2)$ is the two-particle density distribution.

The relationships between the dimensionless probability distributions and the density distributions are given by
\begin{equation}
P_1(p)\propto\rho_1(p),
\end{equation}
\begin{equation}
P_2(p_1,p_2)\propto\rho_2(p_1,p_2).
\end{equation}

In theoretical and modeling studies, the correlation function often uses the ratio
\begin{equation}\label{R2p1p2}
R_2(p_1,p_2)=\frac{\rho_2(p_1,p_2)}{\rho_1(p_1)\rho_1(p_2)},
\end{equation}
which is a six-dimensional representation. However, it is more convenient to use the one-dimensional two-particle correlation function, which is expressed as
\begin{equation}
C_2(Q)=1+\frac{P_2(Q)}{P_{11}(Q)},
\end{equation}
where $Q^2$ is defined as the invariant squared difference of the four-momenta of the two particles:
\begin{equation}\label{q2definition}
Q^2=-(p_1-p_2)^2=(p_1+p_2)^2-4m_{\pi}^2>0.
\end{equation}
The correlation functions are obtained via
\begin{equation}\label{P11Q}
P_{11}(Q)=\int d\phi_1d\phi_2\delta[Q^2+(p_1-p_2)^2]\rho_1(p_1)\rho_1(p_2)
\end{equation}
\begin{equation}\label{P2Q}
P_{2}(Q)=\int d\phi_1d\phi_2\delta[Q^2+(p_1-p_2)^2][\rho_2(p_1,p_2)-\rho_1(p_1)\rho_1(p_2)],
\end{equation}
where the pseudo-correlation in $\rho_2(p_1,p_2)$ is deducted.

\begin{figure}[h]
    \centering
    \includegraphics[width=0.35\textwidth]{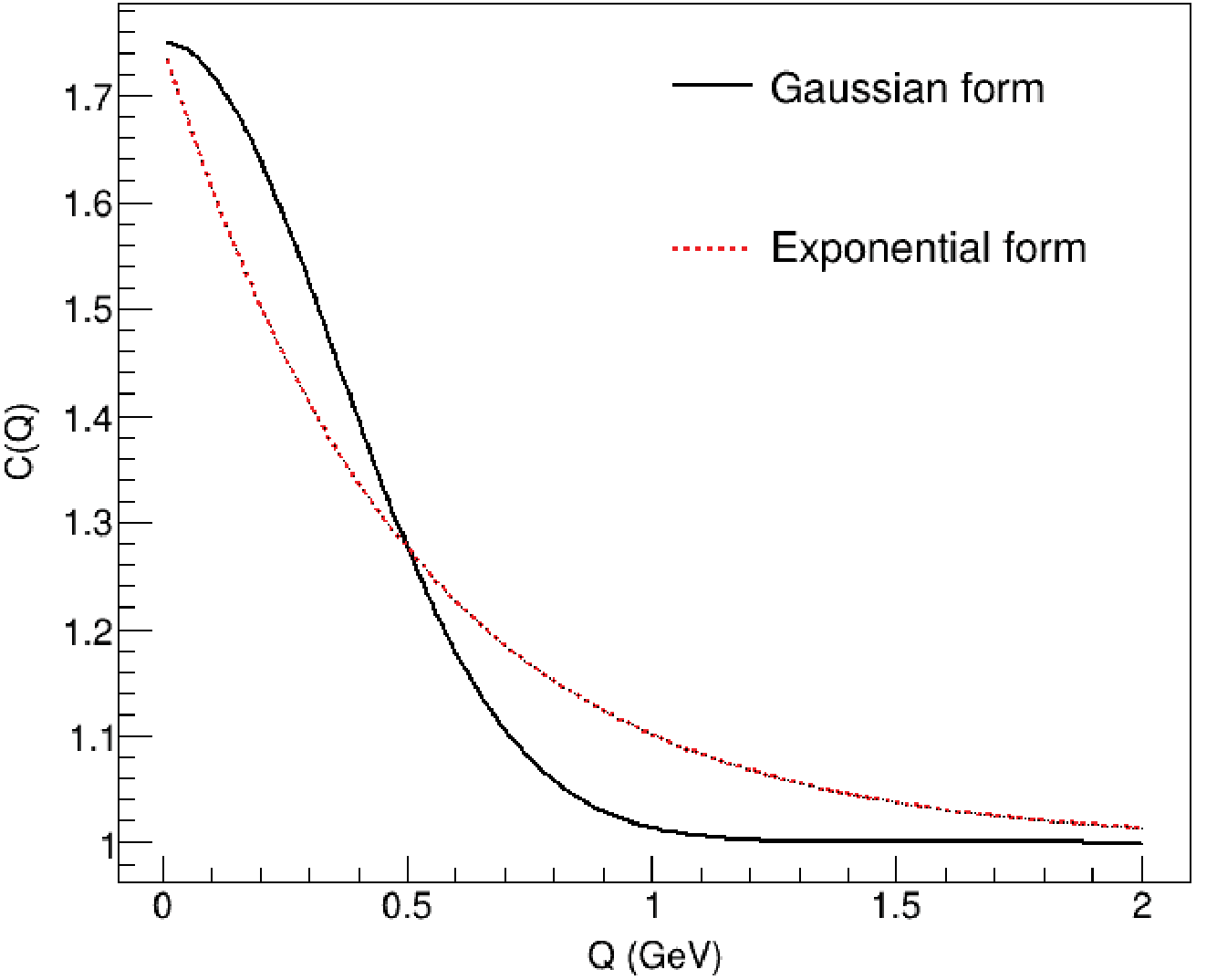}
    \caption{(color online) Line-shapes of the exponential form expressed by Eq.(\ref{LUBOEIexplineshape}) and Gaussian form by Eq.(\ref{LUBOEIgaslineshape}) of BECF.}
    \label{BECexpgaslineshape}
\end{figure}

In experimental studies, BECFs are often parameterized as
\begin{equation}\label{LUBOEIexplineshape}
C(Q)=1+\lambda\exp(-RQ)~~~~~({\rm exponential~form}),
\end{equation}
\begin{equation}\label{LUBOEIgaslineshape}
C(Q)=1+\lambda\exp[-(RQ)^2]~~~~~({\rm Gaussian~form}),
\end{equation}
where $\lambda$ is known as the incoherent or chaotic parameter, and $R$ is often interpreted as the source radius \cite{boal,weiner}. The line-shapes of Eqs.(\ref{LUBOEIexplineshape}) and (\ref{LUBOEIgaslineshape}) are shown in Fig.\ref{BECexpgaslineshape}. In the figure, $\lambda=0.75$ and $R=0.4$ fm are used, but the actual values are determined by fitting the experimental data. According to Eqs. (\ref{LUBOEIexplineshape}) and (\ref{LUBOEIgaslineshape}), the correlation function has two important features: it goes to $1$ at large $Q$ values, and it goes to $1+\lambda$ at $Q=0$.

Because $Q^2$ is defined as a four-dimensional form in Eq.(\ref{q2definition}), the parameter $R$ in Eq.(\ref{LUBOEIexplineshape}) and (\ref{LUBOEIgaslineshape}) does not necessarily reflect the true spatial scale of the hadron source as it also depends on the lifetime of the hadron source. Various models attempt to provide a physical explanation for the parameter $\lambda$ \cite{weiner,boal,bolz,csorgo}. While $\lambda$ may consist of multiple physical factors, the term expressed in Eq.(\ref{lambdaci}) is certainly a component of them.

\section{Correction factors in measurement}

To the shape of the BECF, as measured by the experimental data, can be fit using the correlation density, which is expressed as
\begin{equation}
\rho(Q_i)=K\frac{\Delta n(Q_i)}{\Delta Q_i},
\end{equation}
where $\Delta n(Q_i)$ is the number of particle pairs falling within the interval $(Q_i,Q_i+\Delta Q_i)$, and $K$ is a normalization factor.

The observed density correlation functions are obtained by selecting signal and reference particle pairs from the experimental data. Some preliminary hadrons are unstable and decay into final-state hadrons before reaching the detector. The correlation functions obtained from the decay final-state hadrons are given by
\begin{eqnarray}\label{rhofinsigref}
\rho_{\rm finsig}^{\rm datobs}(Q)&=&K_{\rm finsig}^{\rm datobs}\frac{\Delta n_{\rm finsig}^{\rm datobs}(Q)}{\Delta Q},\\
\rho_{\rm finref}^{\rm datobs}(Q)&=&K_{\rm finref}^{\rm datobs}\frac{\Delta n_{\rm finref}^{\rm datobs}(Q)}{\Delta Q},
\end{eqnarray}
The final state BECF is defined as
\begin{equation}\label{Cfindatobs}
C_{\rm fin}^{\rm datobs}(Q)=\frac{\rho_{\rm finsig}^{\rm datobs}(Q)}{\rho_{\rm finref}^{\rm datobs}(Q)}.
\end{equation}

Ideally, the BECF should be measured using preliminary-state hadrons. If the decayed final states can be traced back to the preliminary states, the preliminary-state density correlation functions are given by
\begin{equation}\label{rhopresigref}
\rho_{\rm presig}^{\rm datobs}(Q)=K_{\rm presig}^{\rm datobs}\frac{\Delta n_{\rm presig}^{\rm datobs}(Q)}{\Delta Q},
\end{equation}
\begin{equation}
\rho_{\rm preref}^{\rm datobs}(Q)=K_{\rm preref}^{\rm datobs}\frac{\Delta n_{\rm preref}^{\rm datobs}(Q)}{\Delta Q},
\end{equation}
and the observed preliminary BECF is
\begin{equation}\label{Cpredatobs}
C_{\rm pre}^{\rm datobs}(Q)=\frac{\rho_{\rm presig}^{\rm datobs}(Q)}{\rho_{\rm preref}^{\rm datobs}(Q)}.
\end{equation}
Both the BECFs expressed by Eqs.(\ref{Cfindatobs}) and (\ref{Cpredatobs}) depend on the data analysis method, and the physical results account for the efficiency correction.

To derive the physical BECF, $C^{\rm phy}(Q)$, from the observed $C_{\rm pre}^{\rm datobs}(Q)$ and $C_{\rm fin}^{\rm datobs}(Q)$, two sets of corrections are required: one that accounts for the efficiency in the analysis-dependent quantities, and another that converts the final-state $C_{\rm fin}(Q)$ into a preliminary-state one $C_{\rm pre}(Q)$.

If MC simulations align well with the experimental data, then the relationship
\begin{equation}
\frac{C^{\rm phy}(Q)}{C^{obs}(Q)}=\frac{C^{\rm simgen}(Q)}{C^{\rm simobs}(Q)},
\end{equation}
holds, where
\begin{equation}
C^{\rm simgen}(Q)=\frac{\rho_{\rm sig}^{\rm simgen}(Q)}{\rho_{\rm ref}^{\rm simgen}(Q)},
\end{equation}
\begin{equation}
C^{\rm simobs}(Q)=\frac{\rho_{\rm sig}^{\rm simobs}(Q)}{\rho_{\rm ref}^{\rm simobs}(Q)}
\end{equation}
are the BECFs obtained by MC simulations at the generator and detector levels, respectively.

For both preliminary- and final-states hadrons simulated at the generator and detector levels, we have:
\begin{equation}\label{cpre}
C_{\rm pre}^{\rm simgen}(Q)=\frac{\rho_{\rm presig}^{\rm simgen}(Q)}{\rho_{\rm preref}^{\rm simgen}(Q)},~
C_{\rm pre}^{\rm simobs}(Q)=\frac{\rho_{\rm presig}^{\rm simobs}(Q)}{\rho_{\rm preref}^{\rm simobs}(Q)},
\end{equation}
\begin{equation}\label{cfin}
C_{\rm fin}^{\rm simgen}(Q)=\frac{\rho_{\rm finsig}^{\rm simgen}(Q)}{\rho_{\rm finref}^{\rm simgen}(Q)},~
C_{\rm fin}^{\rm simobs}(Q)=\frac{\rho_{\rm finsig}^{\rm simobs}(Q)}{\rho_{\rm finref}^{\rm simobs}(Q)}.
\end{equation}

The experimental physics results of the BECFs can be derived from the observed values by applying the efficiency corrections
\begin{equation}
C_{\rm pre}^{\rm datexp}(Q)=\frac{C_{\rm pre}^{\rm datobs}(Q)}{\epsilon_{\rm pre}(Q)},
C_{\rm fin}^{\rm datexp}(Q)=\frac{C_{\rm fin}^{\rm datobs}(Q)}{\epsilon_{\rm fin}(Q)},
\end{equation}
where the efficiencies $\epsilon_{\rm pre}$ and $\epsilon_{\rm fin}$ are obtained via MC simulations:
\begin{equation}
\epsilon_{\rm pre}(Q)=\frac{C_{\rm pre}^{\rm simobs}(Q)}{C_{\rm pre}^{\rm simgen}(Q)},
\epsilon_{\rm fin}(Q)=\frac{C_{\rm fin}^{\rm simobs}(Q)}{C_{\rm fin}^{\rm simgen}(Q)}.
\end{equation}

The relationship of BECFs between the experimental raw data and physics result is
\begin{equation}\label{correctfromfintopre}
C^{\rm phy}(Q)=C_{\rm pre}^{\rm datexp}(Q)=f_{\rm prefin}^{\rm simgen}(Q)\cdot C_{\rm fin}^{\rm datexp}(Q),
\end{equation}
where the recover factor is
\begin{equation}\label{prefinratio}
f_{\rm prefin}^{\rm simgen}(Q)=\frac{C_{\rm pre}^{\rm simgen}(Q)}{C_{\rm fin}^{\rm simgen}(Q)}.
\end{equation}

\section{Monte Carlo simulations}

The MC generators for the Lund model are LUARLW (for the low-to medium-energy region) \cite{bohu} and JETSET (for the high-energy region) \cite{jetset}.

The LUBOEI subroutine in in LUARLW and JETSET employs algorithms based on mean-field potential attraction between identical bosons to simulate the BEC effect. This is achieved by shifting the final-state momenta of identical bosons closer together while maintaining energy-momentum conservation within the event \cite{EurPhysJC21998165,plb3511995293,BECinMP}. LUBOEI offers two options for simulating the BEC effect: the exponential form (given in Eq.(\ref{LUBOEIexplineshape})) and the Gaussian form (given in Eq.(\ref{LUBOEIgaslineshape})). An analysis of the BESIII data indicated that the Gaussian form agreed well with the experimental data; thus, it was used in the MC simulations.

In this section, we present various BECF line shapes simulated at the generator level. The signal samples typically include $\pi^+\pi^+$, $\pi^-\pi^-$, and $\pi^0\pi^0$, while the reference samples include $\pi^+\pi^-$ and $\pi^\pm\pi^0$. We provide comparisons between the preliminary and final states for different signal and reference samples at the generator level.

\begin{figure}[htb]
    \centering
    \includegraphics[width=0.35\textwidth]{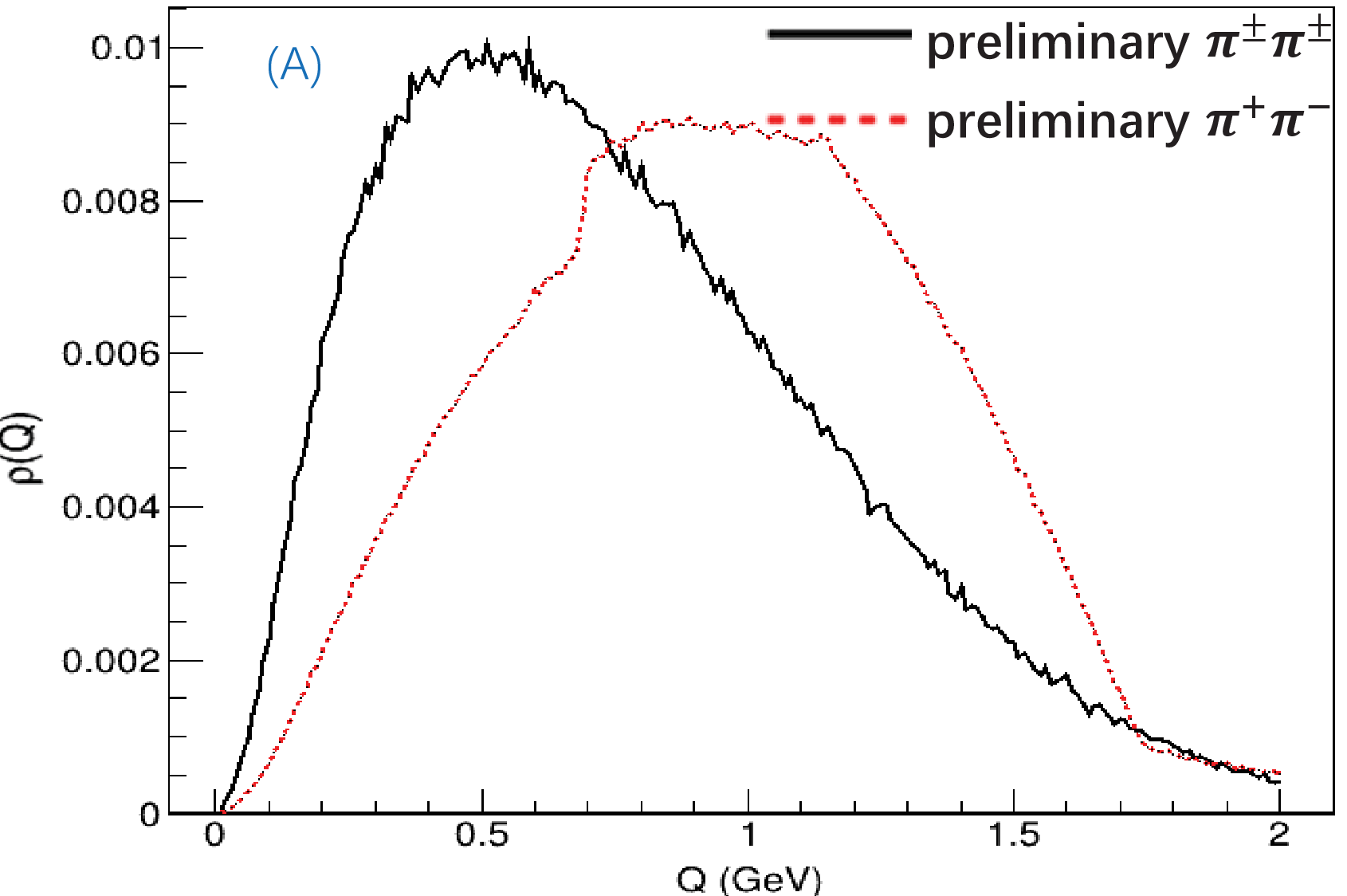}
    \includegraphics[width=0.35\textwidth]{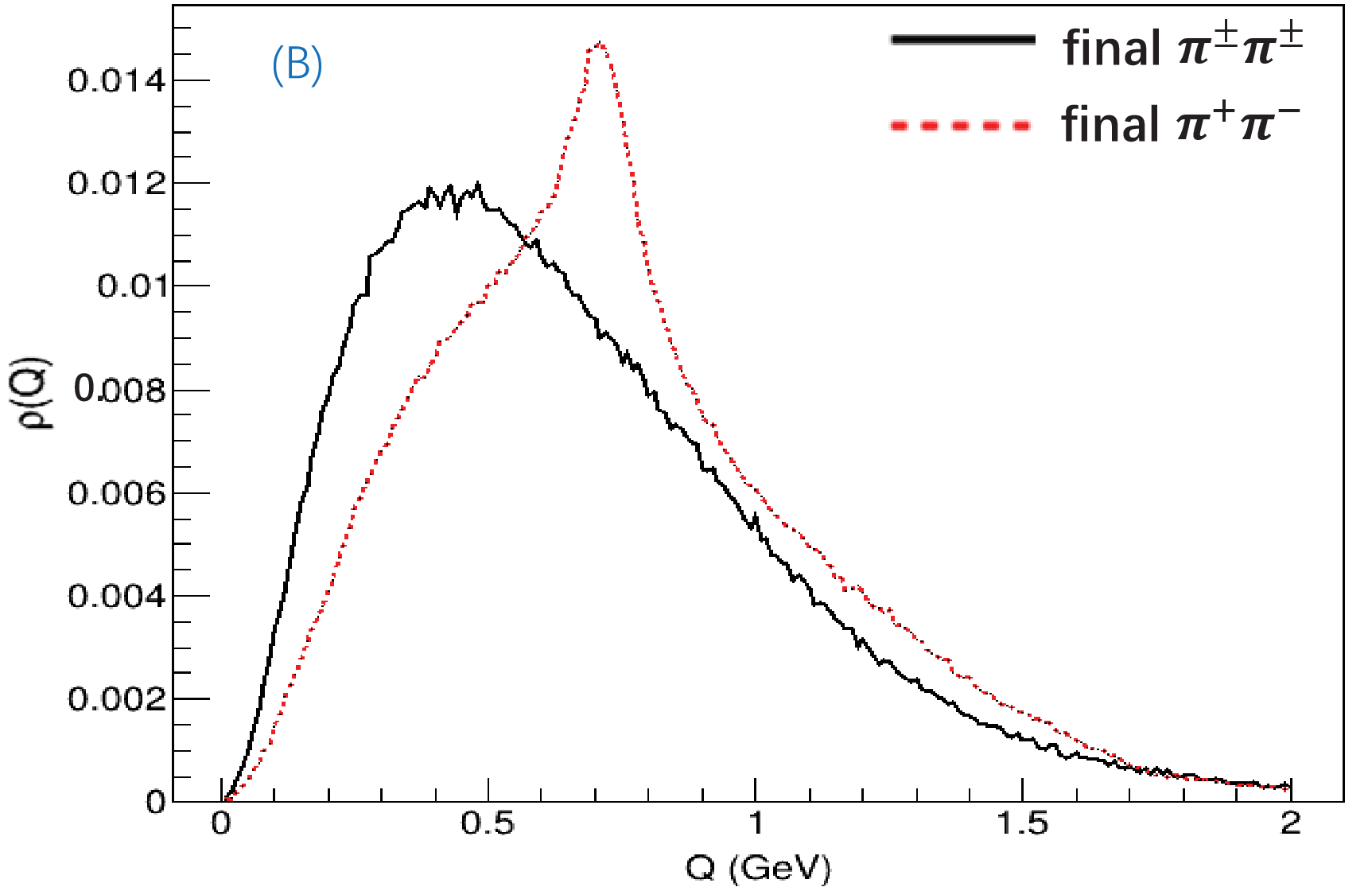}
    \caption{(color online) Density correlation function $\rho(Q)$ of the preliminary states (A) and final states (B) of the signal and reference samples at 3 GeV.}
    \label{rhosigrefprefin}
\end{figure}

In the MC simulations, we could generate either preliminary hadron states only or allow unstable hadrons to decay into final states. The inclusive line shapes of the normalized density distributions $\rho_{\rm presig}^{\rm simgen}(Q)$ and $\rho_{\rm preref}^{\rm simgen}(Q)$ in Eq.(\ref{cpre}) and those of $\rho_{\rm finsig}^{\rm simgen}(Q)$ and $\rho_{\rm finref}^{\rm simgen}(Q)$ in Eq.(\ref{cfin}) are shown in Fig.\ref{rhosigrefprefin}. The line shapes of $\rho_{\rm presig}^{\rm simgen}(Q)$ and $\rho_{\rm finsig}^{\rm simgen}(Q)$ tend towards the low $Q$ end, which is a characteristic feature of the BEC effect.

In the reference sample, some $\pi^+\pi^-$ pairs are the decay final states of resonances such as $\rho$, $\eta$, $\eta'$, $\omega$ and $\phi$. In Fig.\ref{rhosigrefprefin}(B), the convexity of $\rho_{\rm finref}^{\rm simgen}(Q)$ is caused by resonant effects. We can exclude these $\pi^+\pi^-$ pairs by requiring their squared invariant mass $q^2=(p_{\pi^+}+p_{\pi^-})^2$ to lie outside the resonance peak $[(M_R-\Gamma)^2,(M_R+\Gamma)^2]$. However, this requirement inevitably removes some $\pi^+\pi^-$ pairs from non-resonant states.

Various schemes for selecting reference samples have been investigated by different collaborations \cite{eurphysjc822022608}. However, these schemes can violate energy-momentum conservation. As noted in Ref.\cite{boal}, the correlations arising from energy-momentum conservation are stronger than those arising from the BEC effect. It is expected that the biases introduced by these schemes can be corrected using high-quality MC simulations and the correction factor given in Eq.(\ref{prefinratio}).

\begin{figure}[htb]
    \centering
    \includegraphics[width=0.35\textwidth]{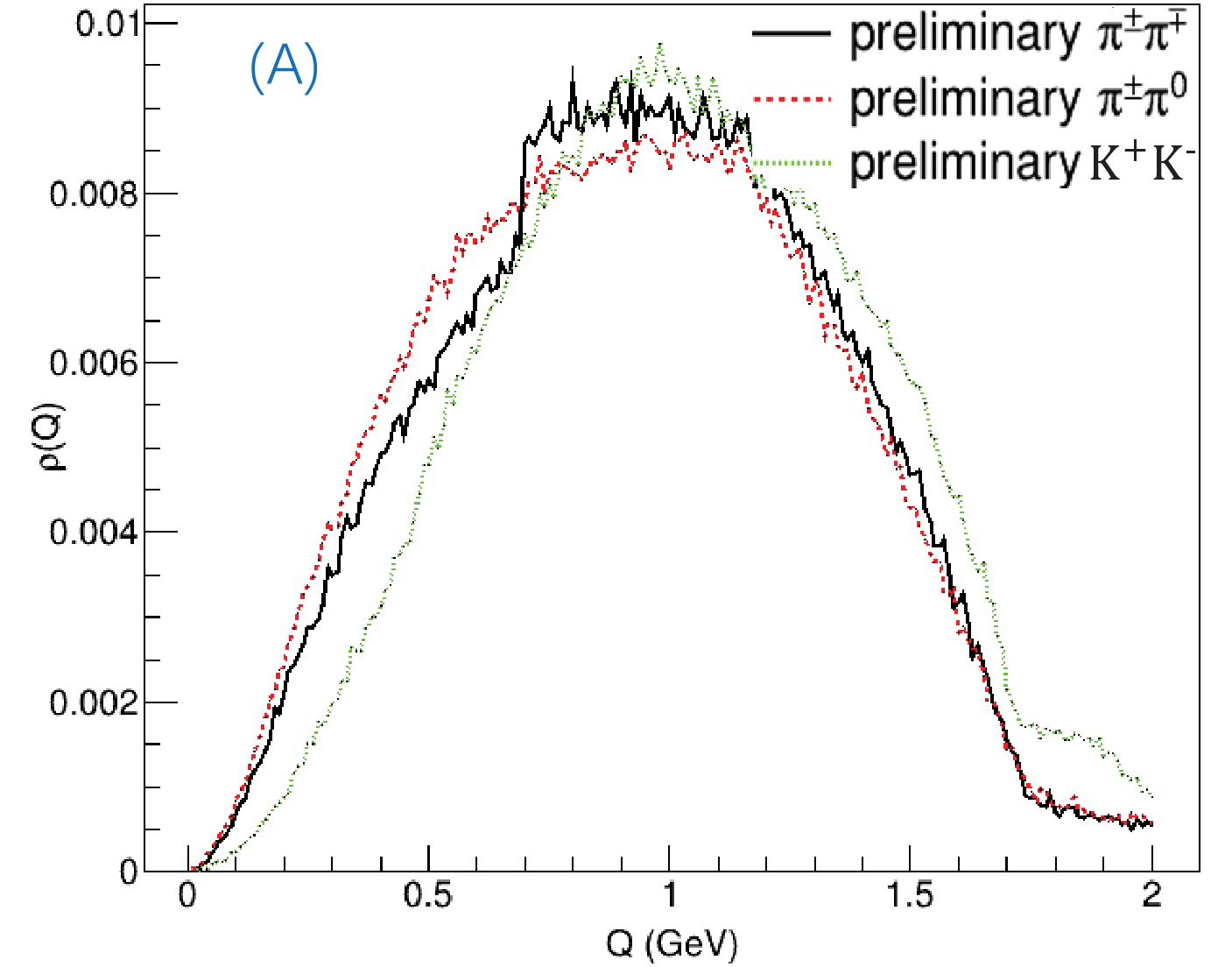}
    \includegraphics[width=0.35\textwidth]{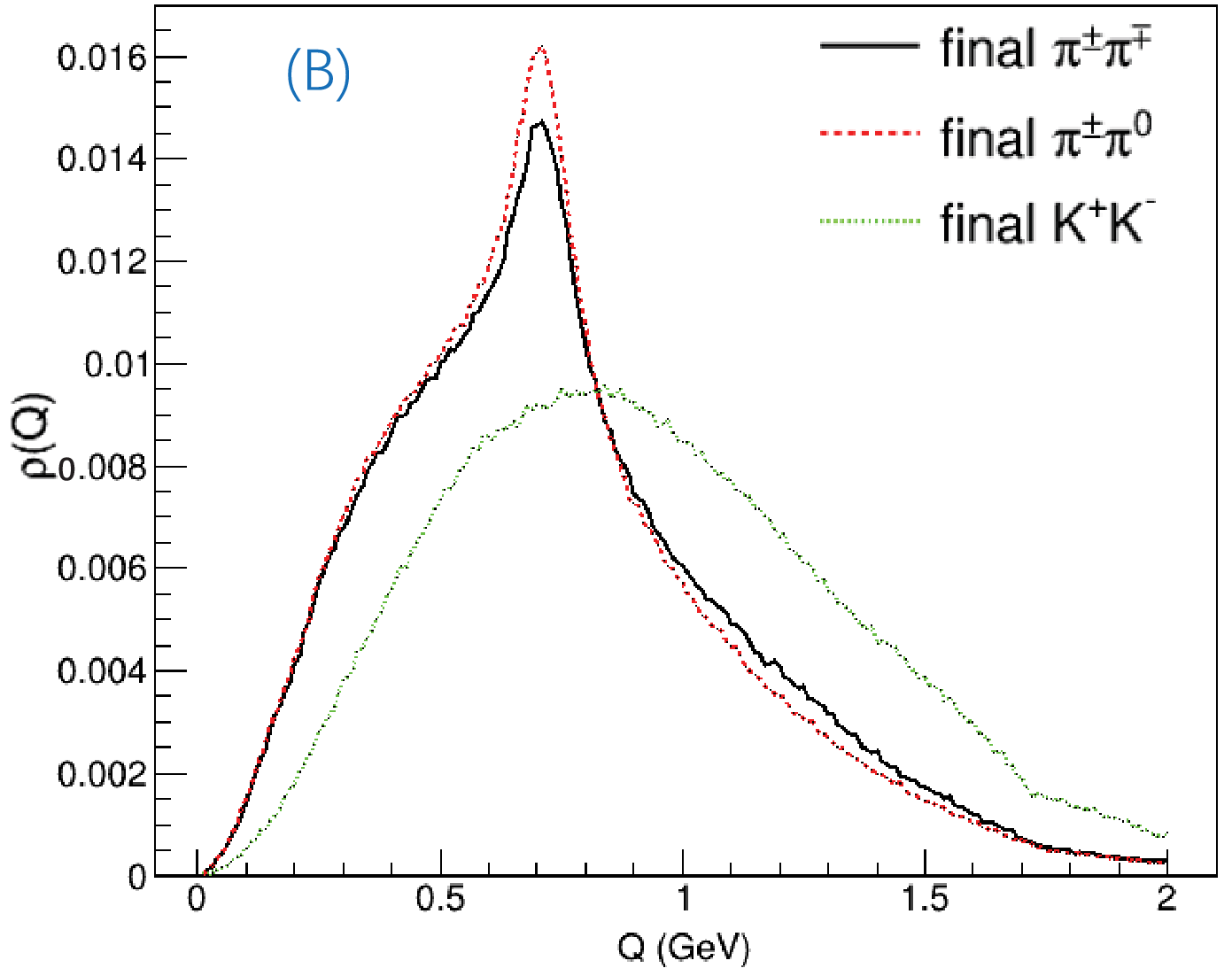}
    \caption{(color online) Density correlation function of the preliminary and final states for the reference samples $\pi^+\pi^-$, $\pi^{\pm}\pi^0$ and $K^+K^-$ at 3 GeV.}
    \label{rhoprefinpippimpicpi0diff}
\end{figure}

Figure \ref{rhoprefinpippimpicpi0diff} shows the density correlation functions $\rho(Q)$ of the preliminary and final states for $\pi^+\pi^-$, $\pi^{\pm}\pi^0$ and $K^+K^-$. If $\rho(Q)$ consists of only $\pi^+\pi^-$ pairs or all of the $\pi^+\pi^-$, $\pi^{\pm}\pi^0$ and $K^+K^-$ pairs, the line shape of $C(Q)$ differs.

\begin{figure}[htb]
    \centering
    \includegraphics[width=0.35\textwidth]{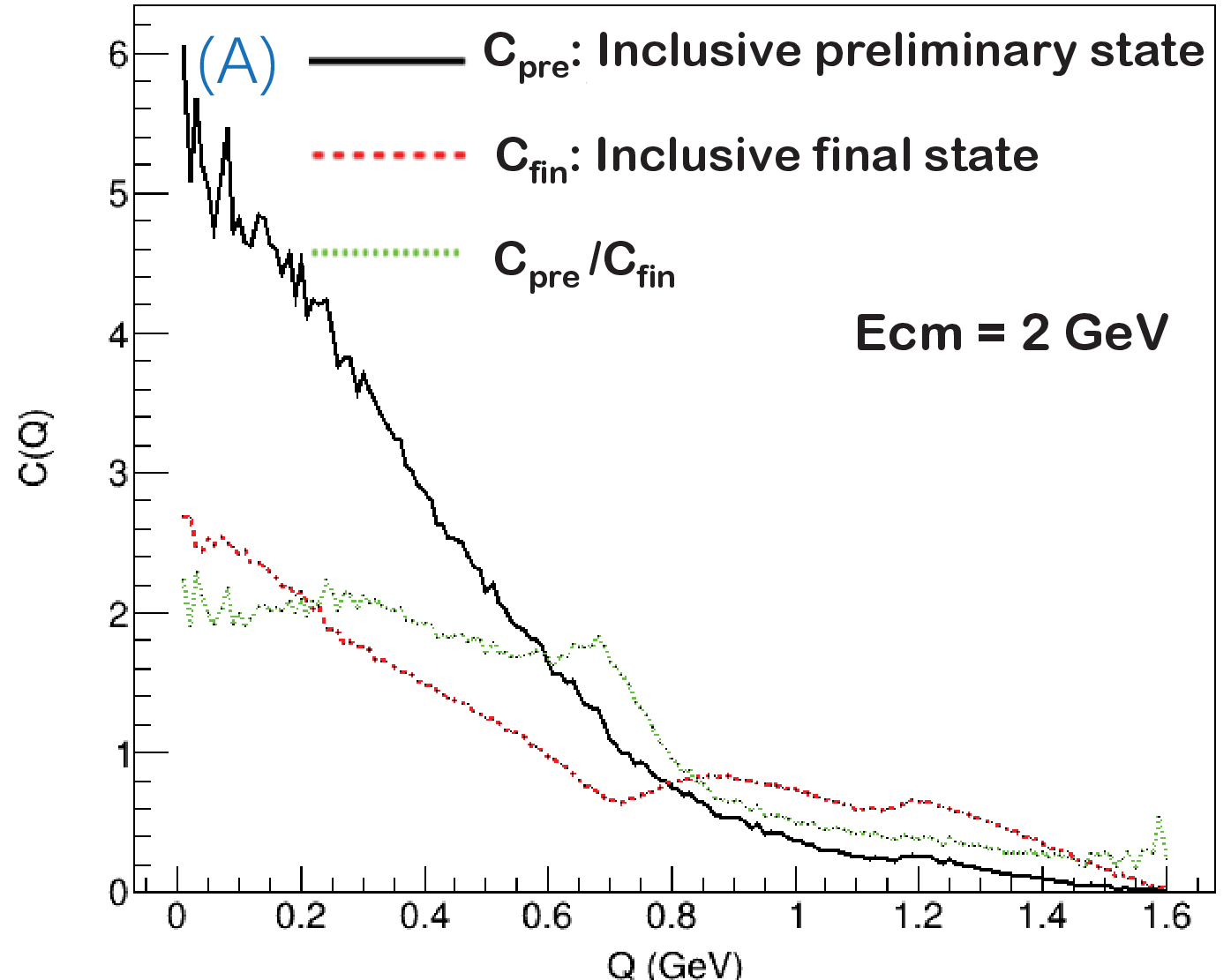}
    \includegraphics[width=0.35\textwidth]{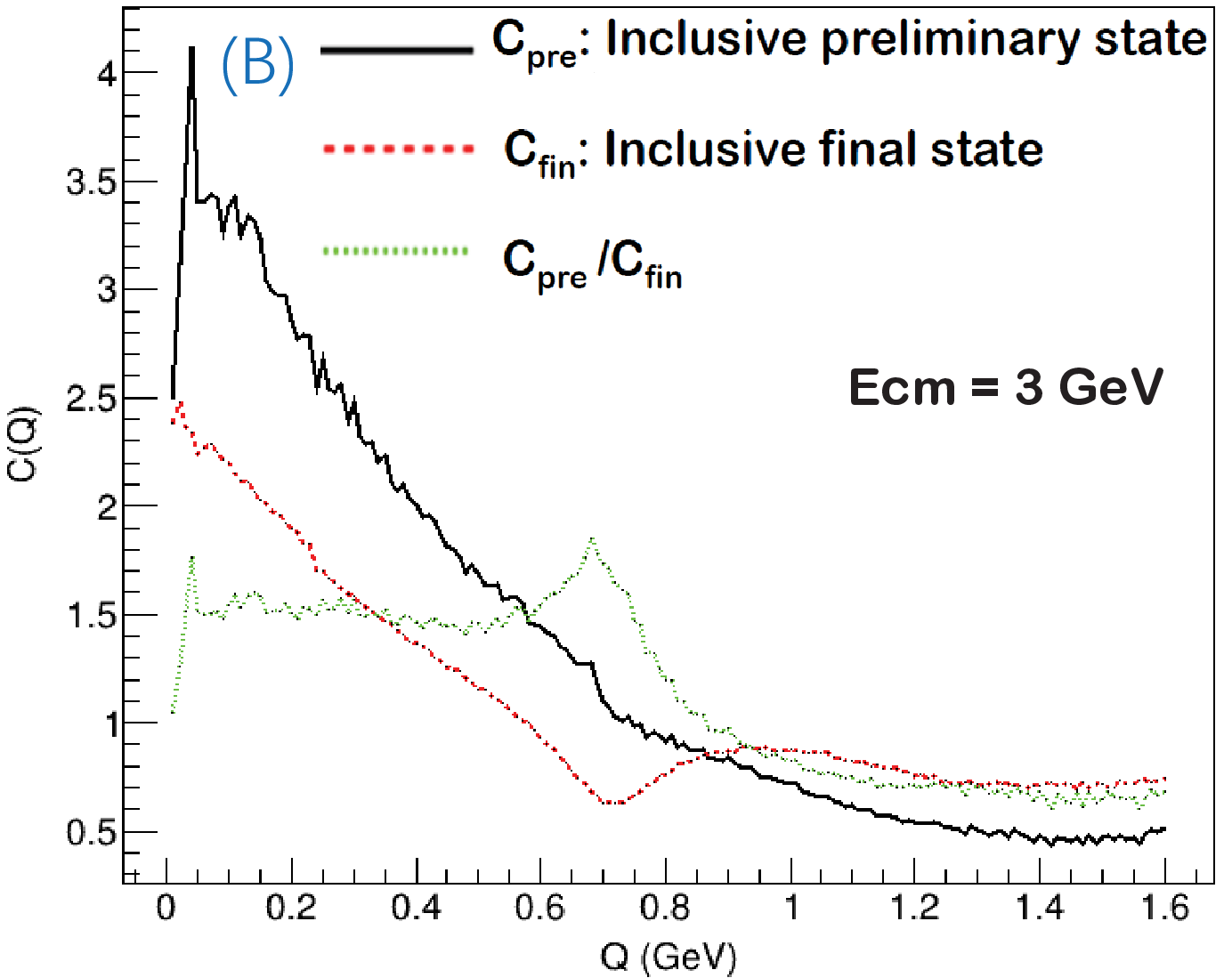}
    \caption{(color online) BECF of the inclusive preliminary and final states and the corresponding ratio at 2 GeV(top) and 3 GeV(bottom).}
    \label{cprefinat2and3gev}
\end{figure}

Figure \ref{cprefinat2and3gev} illustrates the BECF of the inclusive preliminary and final states, along with the corresponding ratio as presented in Eq.(\ref{prefinratio}) at 2 GeV and 3 GeV. The figures indicate that the line shapes of $C(Q)$ are energy-dependent within the BEPCII energy range. Figure \ref{cexluprefin} shows the dependence of the BECF preliminary and final states for the semi-exclusive continuous states at 3 GeV.

\begin{figure}[htb]
    \centering
    \includegraphics[width=0.35\textwidth]{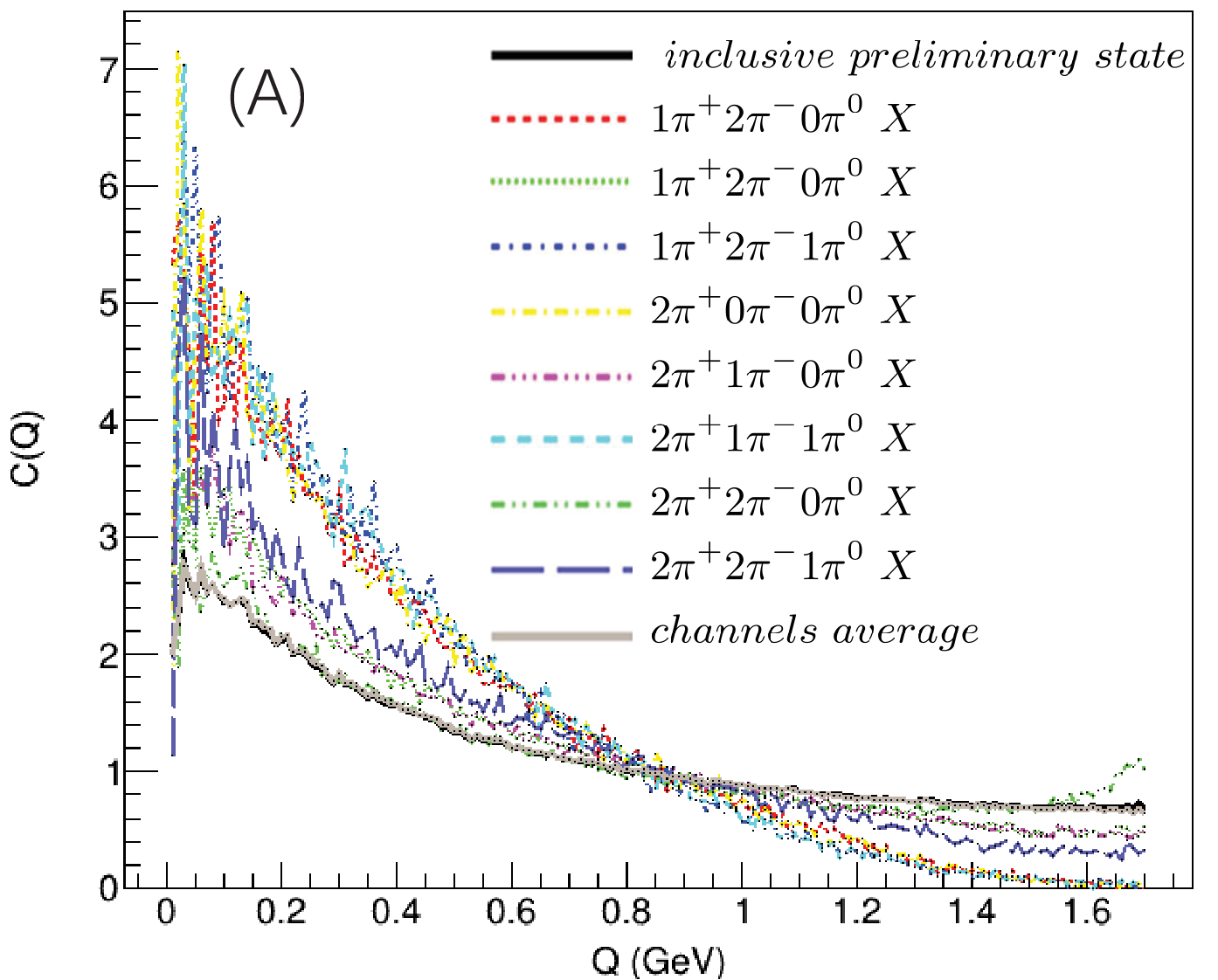}
    \includegraphics[width=0.35\textwidth]{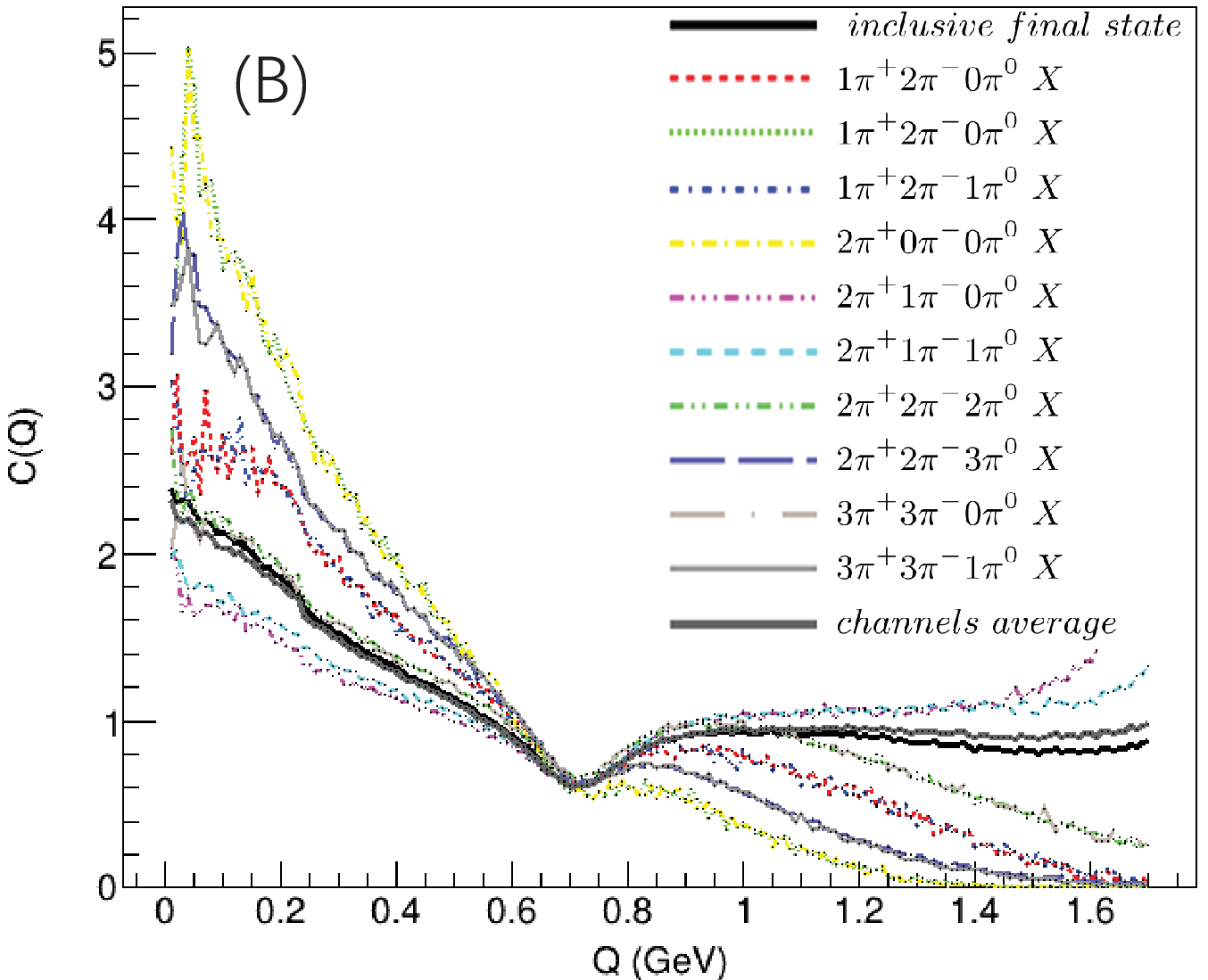}
    \caption{(color online) BECF of the semi-exclusive preliminary and secondary decay final states of the three-gluon decay mode of $J/\psi$.}
    \label{cJpsiexluprefin}
\end{figure}

The $J/\psi$ particle is produced via $e^+e^-\to\gamma^\ast\to c\bar{c}\to J/\psi$, and it decays to preliminary states and/or stable final states. It is interesting to examine whether the BEC effect manifests in the preliminary and secondary final states of $J/\psi$ decay. MC simulations indicate that both preliminary and secondary decay final states exhibit the BEC effect, as shown in Fig.\ref{cJpsiexluprefin}, which shows that the line shapes of $C(Q)$ and $J/\psi$ similar to those of the continuous states.

\begin{figure}[htb]
    \centering
    \includegraphics[width=0.35\textwidth]{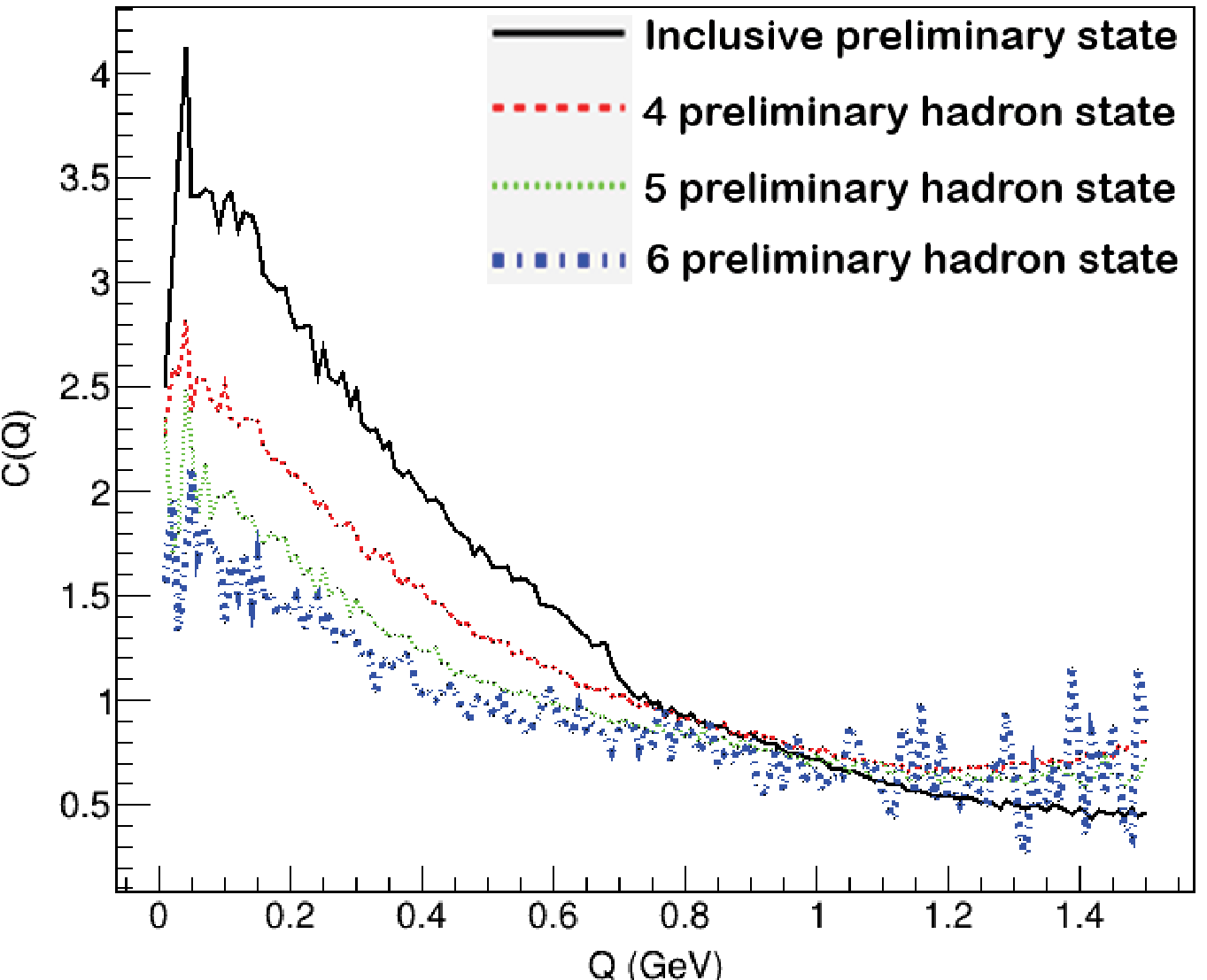}
    \includegraphics[width=0.35\textwidth]{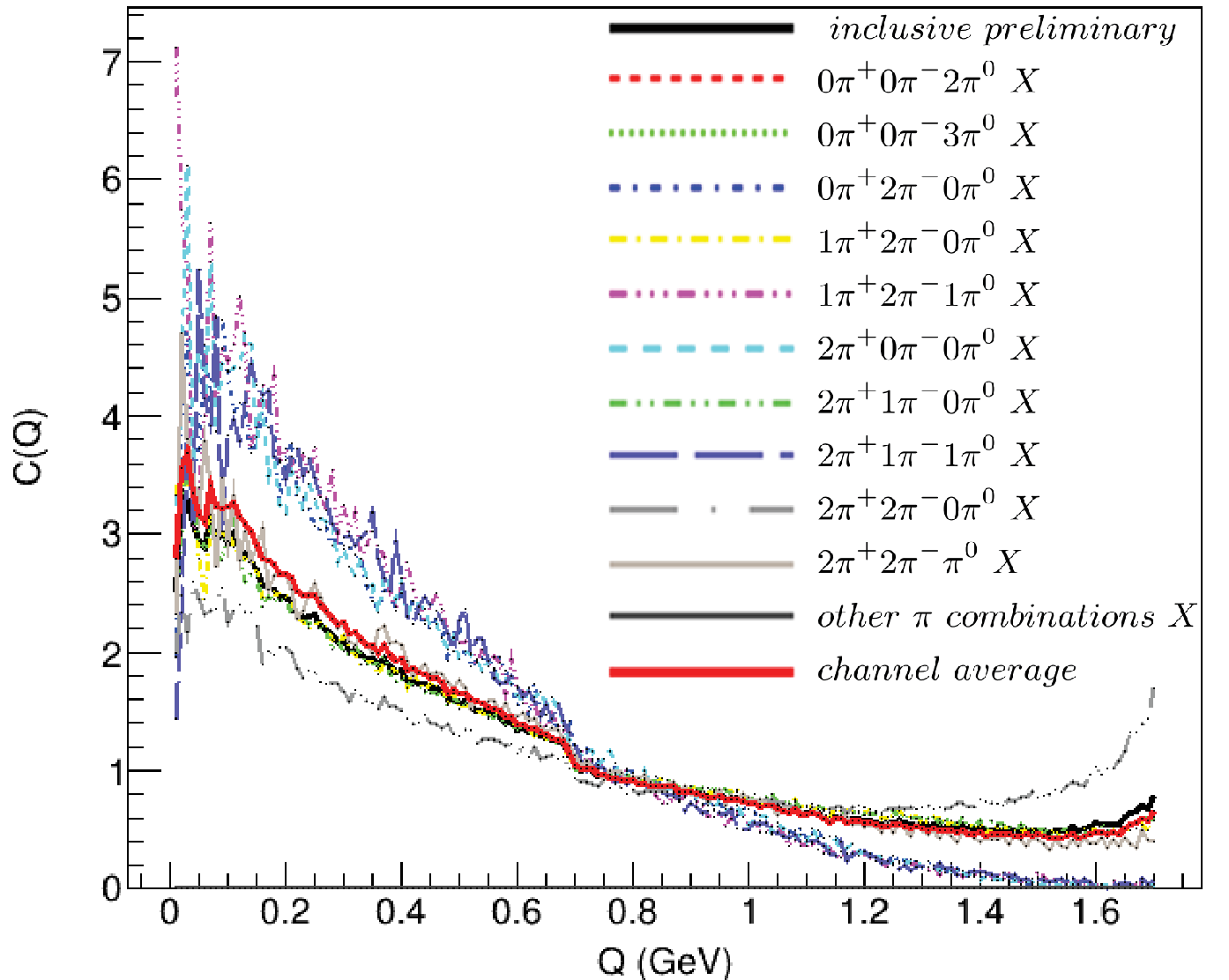}
    \caption{(color online) Left: BECF of inclusive hadron states with different preliminary multiplicities. Right: BECF of a selected of  semiexclusive channels. The reference sample include $\pi^+\pi^-$ and $\pi^\pm\pi^0$.}
    \label{cpre456hadgassemiinclu}
\end{figure}

The Lund model predicts that the hadron emission amplitude with a fixed multiplicity given by Eq.(\ref{fnxipimc}) differs from that of the inclusive case given by Eq.(\ref{fxipimc}). The left figure of Figs.\ref{cpre456hadgassemiinclu} shows the BECF for preliminary hadron states with $n=4,5$ and $6$, as well as for inclusive states, and the right one demonstrates that the line shape of the BECF depends on the choice of signal and reference samples. This implies that the BECF of the hadron source should be measured exclusively, as inclusive measurements can only capture the average features of the hadron source.

\begin{figure}[htb]
    \centering
    \includegraphics[width=0.35\textwidth]{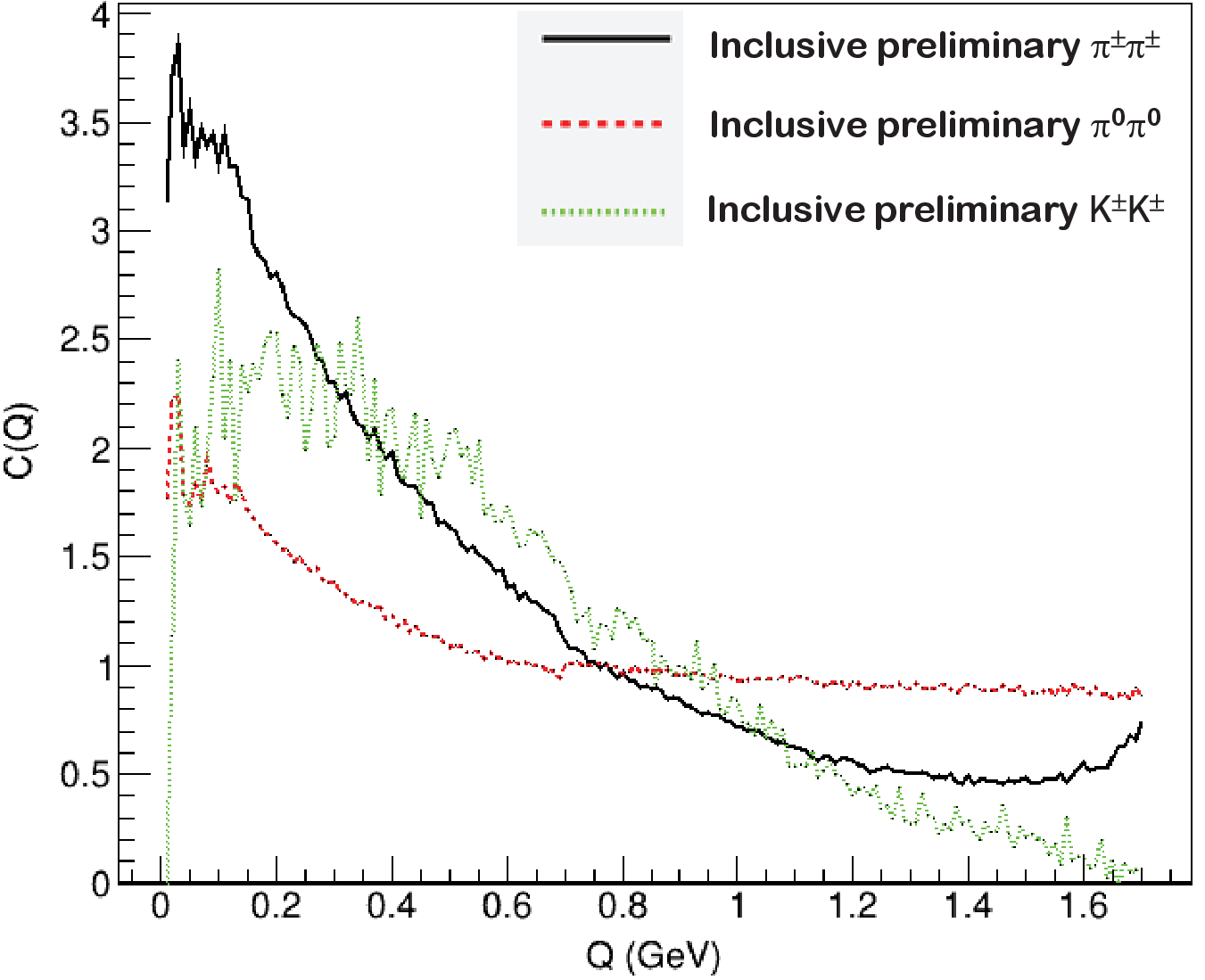}
    \includegraphics[width=0.35\textwidth]{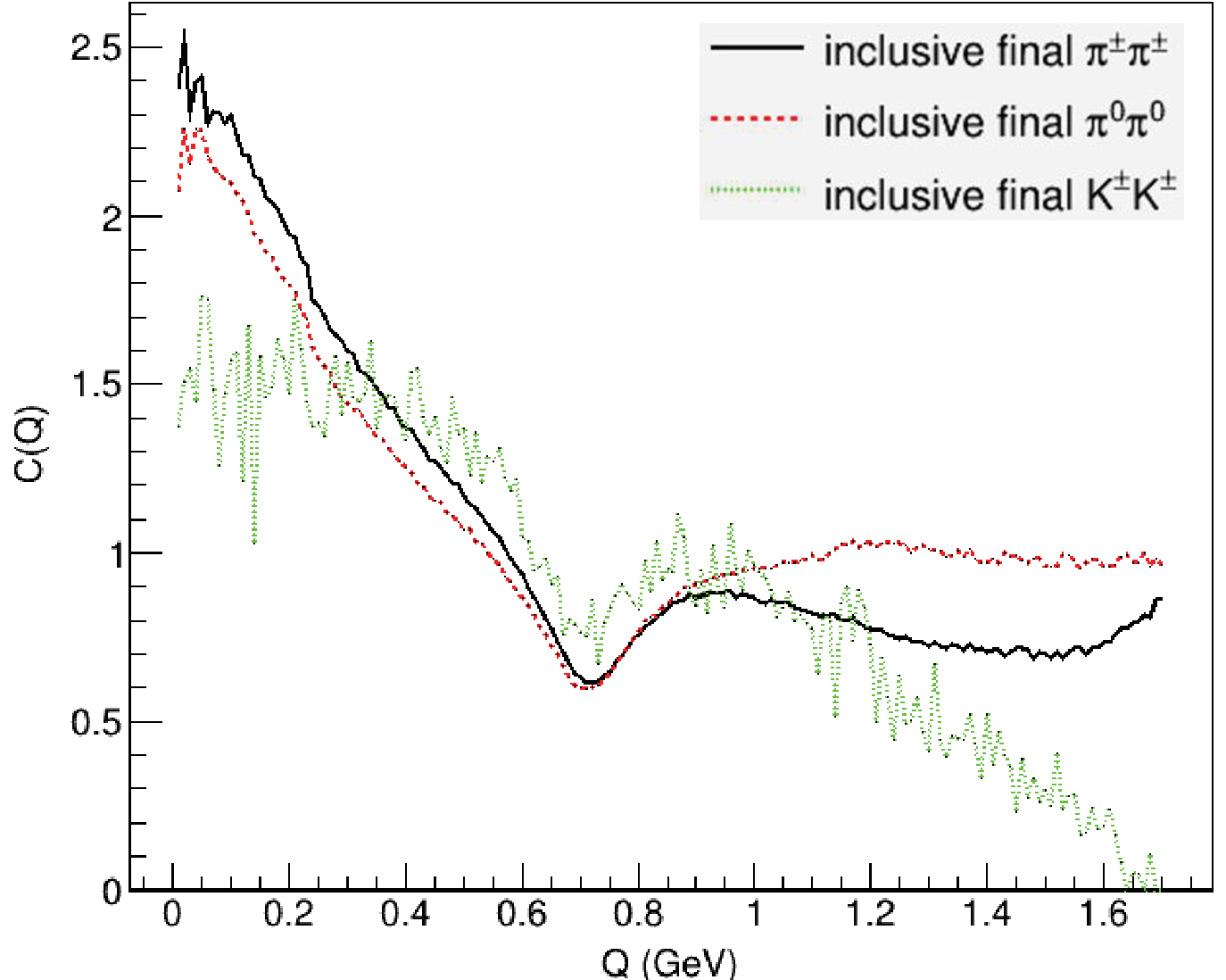}
    \caption{(color online) BECF of the preliminary and final states of $\pi^\pm\pi^\pm$, $\pi^0\pi^0$ and $K^{\pm}K^{\pm}$.}
    \label{Cprefinpicpi0kkgas}
\end{figure}

\begin{figure}[htb]
    \centering
    \includegraphics[width=0.35\textwidth]{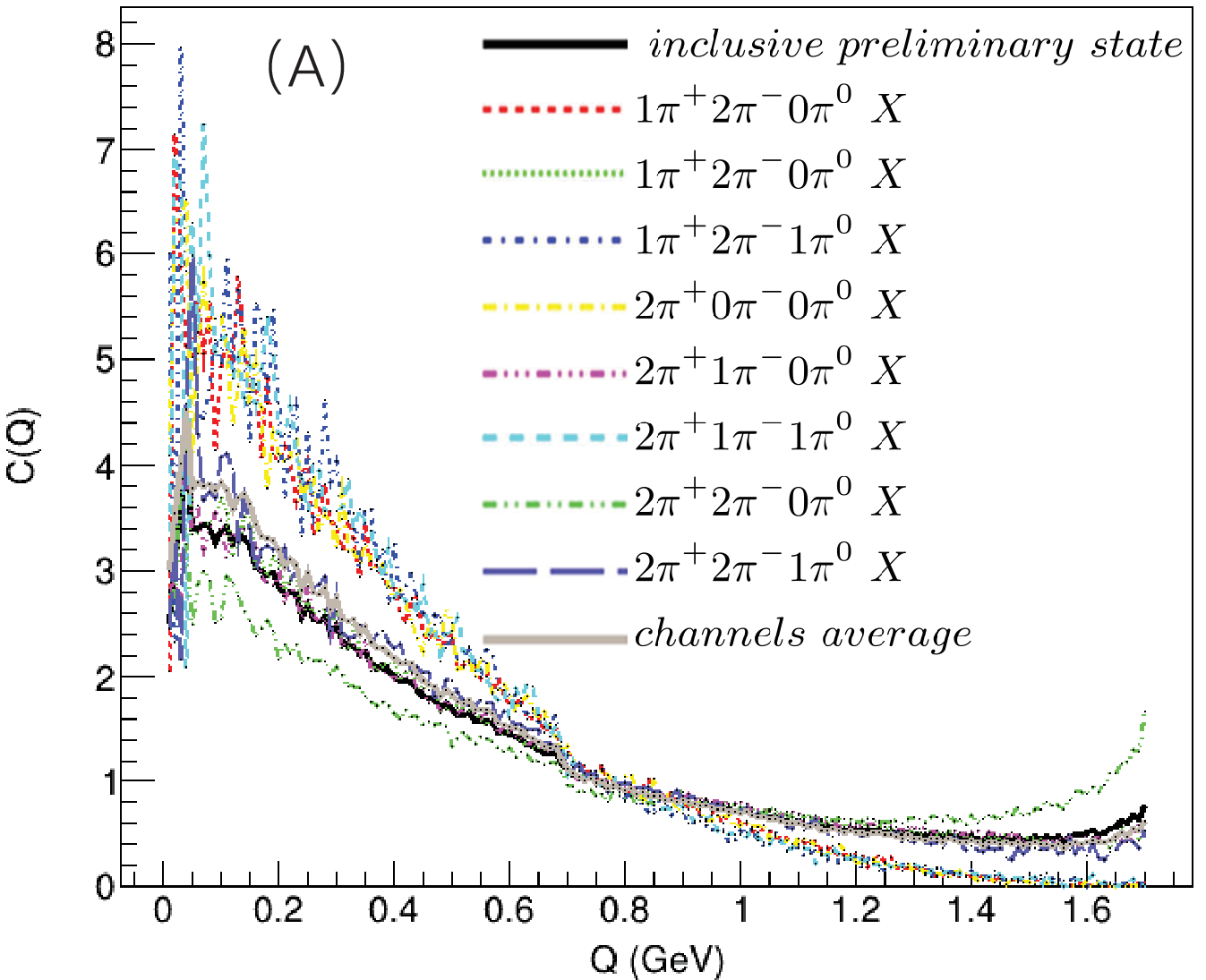}
    \includegraphics[width=0.35\textwidth]{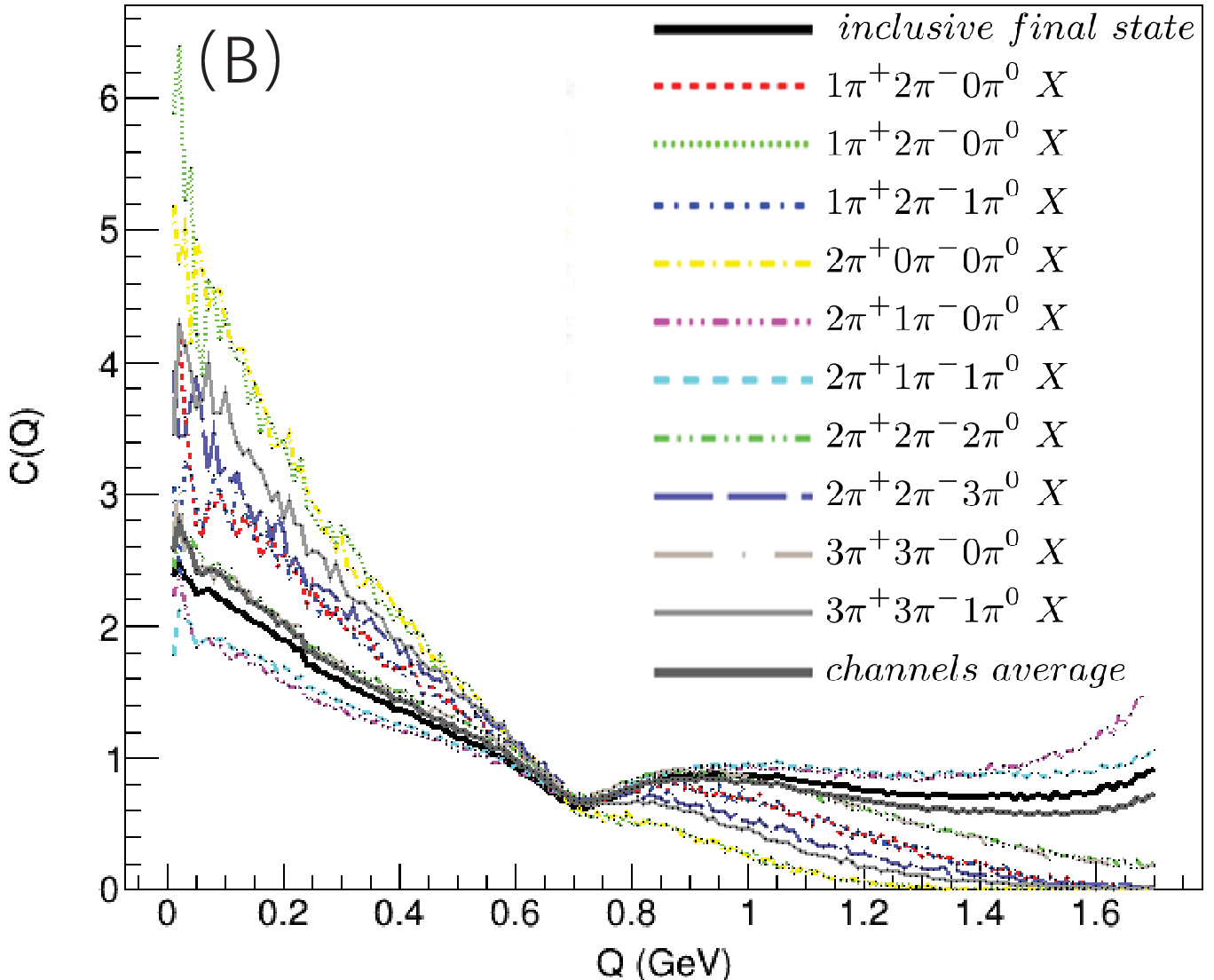}
    \caption{(color online) The BECF of semi-exclusive continuous preliminary production and final states.}
    \label{cexluprefin}
\end{figure}

Figure \ref{Cprefinpicpi0kkgas} shows the BECF for the preliminary and final states of $\pi^\pm\pi^\pm$, $\pi^0\pi^0$ and $K^{\pm}K^{\pm}$, $\pi^\pm\pi^\pm$, $\pi^0\pi^0$ and $K^{\pm}K^{\pm}$, indicating that the BECF depends on the choice of signal samples.

Eqs.(\ref{LUBOEIexplineshape}) and (\ref{LUBOEIgaslineshape}) are simplified parametric forms that describe the main characteristics of the BECF. In the MC simulations, LUBOEI achieves the BEC effect via the mean-field method, and the actual effects also depend on the specific hadron states. Compared with Eqs.(\ref{LUBOEIexplineshape}) and (\ref{LUBOEIgaslineshape}), the practical BECFs are multiplied by the corresponding normalization factors, as is done in most experimental data analyses \cite{opal,na22}, and thus the actual distributions shown in Figs.\ref{cprefinat2and3gev}-\ref{Cprefinpicpi0kkgas} appear different to those shown in Fig.\ref{BECexpgaslineshape}, but the overall trends are similar.

\begin{figure}[htb]
    \centering
    \includegraphics[width=0.35\textwidth]{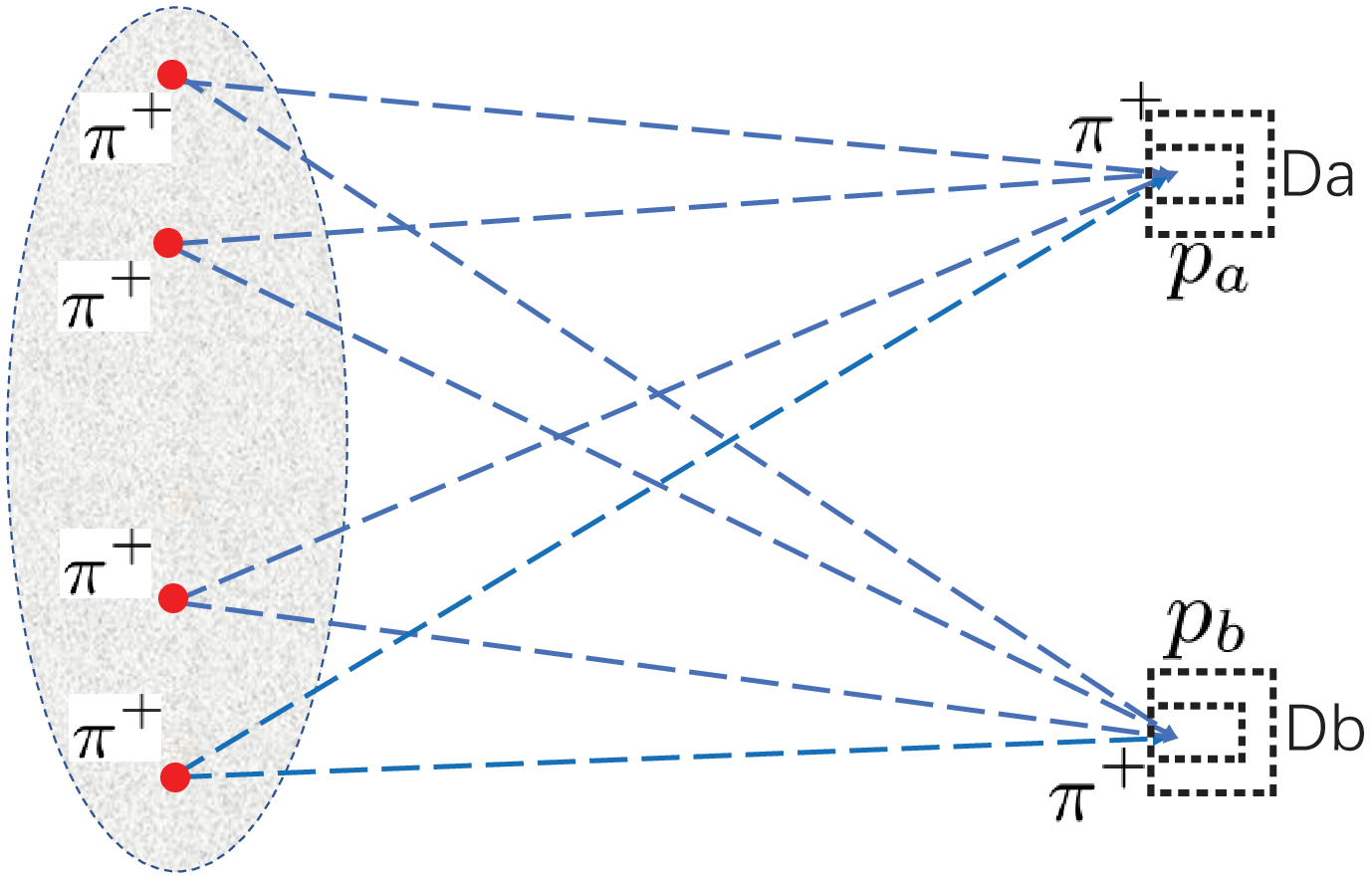}
    \includegraphics[width=0.35\textwidth]{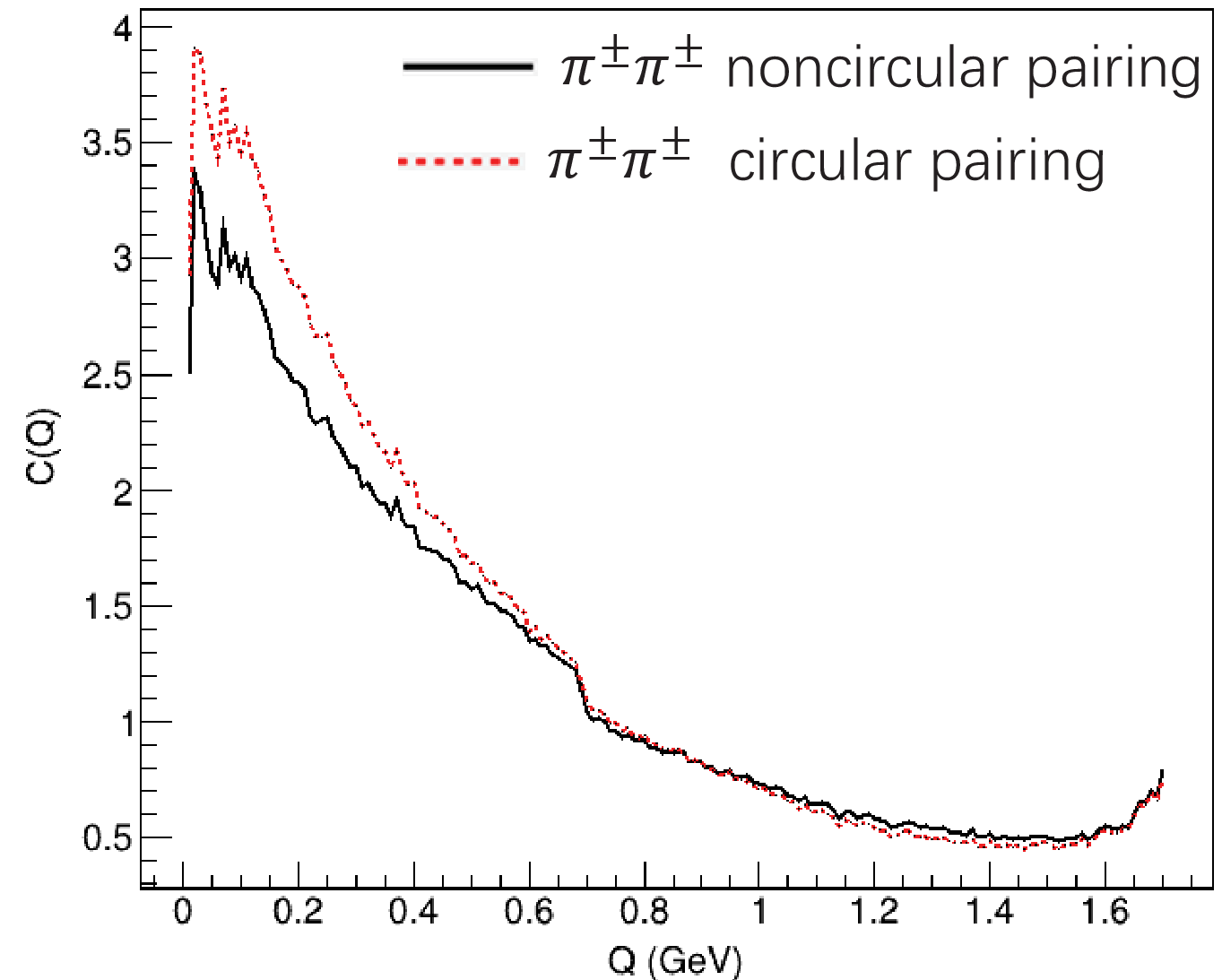}
    \caption{(color online) Left: Four identical $\pi$ productions detected by Da and Db. Right:  Comparison between the BECF line-shapes for circular and noncircular combination of $\pi^\pm\pi^\pm$.}
    \label{hadronscoucepipipipi}
\end{figure}

In previous experiments, signal samples were selected using cyclic combinations, in which any boson could be combined with other identical ones multiple times. This method results in non-independent $Q$ values, introducing extra correlations into $\rho(Q)$. The left figure of Figs.\ref{hadronscoucepipipipi} shows that an event with four identical pions can have six possible cyclic combinations, and the right figure of Figs.\ref{hadronscoucepipipipi} illustrates the line shapes of $C(Q)$ obtained for circular and non-circular combinations of pion pairs. The former introduces extra correlation at small $Q$ value, which is not due to the BEC effect. The correct approach to selecting signal samples is the latter, but the number of identical signal pions is lower.

An alternative generator HYBRID \cite{hybrid} was also used in this study, and it obtained similar results. It combines PHOKARA \cite{phokara}, CONEXC \cite{conexc} and LUARLW, which were developed for the $R$ value measurement at BESIII \cite{besiiir}.

\section{Summary and prospect}

The semi-classical Lund string fragmentation model is currently the only phenomenological model that can capture an accurate physical picture of the hadron production source and correlate the spatiotemporal distribution of the hadron source with the energy-momentum distribution of the produced hadrons.

Starting with the Lund model, in this study we explored basic concepts and formulas related to the BEC effect and performed MC simulations. Through these feasibility studies, several properties of the BEC effect were identified:

(1).~The Lund model establishes the relationships between the space-time points of the hadron source and the momentum of the emitted hadron.

(2).~In addition to the coherent or incoherent components of the hadron source, the status of the particle detector also affects the measured values of the incoherence parameter $\lambda$ and the source scale $R$.

(3).~Different choices of signal and reference samples result in different line shapes for the BECF.

(4).~Experimental measurements of the BEC effect should be conducted exclusively.

(5).~The space-time distribution characteristics of the hadron source are energy-dependent.

(6).~We demonstrated how to convert the observed final state BECF into the preliminary state one using MC simulations.

\begin{center}
\begin{tabular}{ccc}\hline\hline
~$E_{cm} (GeV)$~& ~${\cal L}_{int.}~({\rm pb}^{-1})$~ & ~$N_{had}~(10^6)$~ \\\hline
2.125           & 108.5                         &  5.89  \\
2.396           & ~66.9                         &  2.85  \\
2.645           & ~67.8                         &  2.35  \\
2.900           & 106.0                         &  3.06  \\
3.080           & 126.0                         &  3.09  \\\hline
J/$\psi$        &                               &  $10^4$  \\
$\psi$(2S)      &                               &  3$\times10^3$\\
\hline\hline
\end{tabular}\\
\vspace{1.5mm}
Table 2. Data samples collected at BESIII.
\end{center}
Table 2 presents the integrated luminosity and number of hadron events of the data samples collected using large-sample statistics at BESIII, which enable precise measurements of the BECF at these energies.

The measurement of the BECF using the data collected at BESIII is promising. The pion pairs $\pi^\pm\pi^\pm$ and $\pi^+\pi^-$ were selected as the signal and reference samples, respectively. The generator LUARLW was used to simulate the hadron production, while GEANT4 \cite{GEANT4} was used to simulate the propagation of the particles and their interaction with the detector. The primary systematic uncertainty originated from the hadron generators, for which the difference between LUARLW and HYBRID was taken as an estimation.

A preliminary analysis of the BESIII data indicated that several million same-sign charged pion pairs are available at the five continuum energies. It is expected that the measurement precision for the hadron source parameters, obtained by fitting the line shape of $C(Q)$, will be better than $6\%$ for $\lambda$ and better than $3\%$ for $R$. In the future, data collected for the $J/\psi$ and $\psi(2S)$ peaks will be utilized to measure the BEC effect, further enhancing our understanding of the hadron source.
\bigskip

\begin{acknowledgments}
The authors thank Mr. Wilson J. Huang for proofreading the manuscript. This work is supported by National Natural Science Foundation of China under Contracts No. 11275211,  11335008 and 12035013, National Key R$\&$D Propram of China under Contract No. 2020YFA0406403.
\end{acknowledgments}



\begin{thebibliography}{90}

\bibitem{lundmodel}
B. Andersson, {\it The Lund Model}, (Cambridge University Press, 1998)p. 158

\bibitem{bohu}
B. Andersson and H. Hu, hep-ph/9910285

\bibitem{ioffe}
B.L. Ioffe et al., {\it Hard Processes} Volume 1, (Noth-Holland Physics Publising, 1984)

\bibitem{boal}
D.H. Boal and C.K. Gelbke, Rev. Mod. Phys. {\bf 62}, 553 (1990)

\bibitem{weiner}
R.M. Weiner, Phys. Rept. {\bf 327}, 249 (2000)

\bibitem{csorgo}
T. Cs$\ddot{\rm o}$rg$\ddot{\rm o}$ et al., Eur. Phys. J. C {\bf 9}, 275 (1999),
T. Cs$\ddot{\rm o}$rg$\ddot{\rm o}$ et al., Eur. Phys. J. C {\bf 36}, 67 (2004)

\bibitem{phenix}
T. Nov$\acute{a}$k et al., (PHENIX Collaboration), arXiv:2304.09580 [nucl-ex]

\bibitem{adare}
A. Adare et al., Phys. Rev. C {\bf 97}, 064911 (2018)6

\bibitem{boutemeur}
M. Boutemeur, Nucl. Phys. B(Proc. Suppl.) {\bf 121}, 82 (2003)

\bibitem{markii}
I. Juricic et al., (MARKII Collaboration), Phys. Rev. D {\bf 39}, 1 (1989)

\bibitem{amy}
S. K. Choi et al., (AMY Collaboration), Phys. Lett. B {\bf 355}, 406 (1995)

\bibitem{opal}
G. Alexander et al., (OPAL Collaboration), Z. Phys. C {\bf 72}, 389 (1996)

\bibitem{l3}
P. Achard et al., (L3 Collaboration), Phys. Lett. B {\bf 524}, 55 (2002)

\bibitem{achard}
P. Achard et al., (L3 Collaboration), Eur. Phys. J. C {\bf 71}, 1648 (2011)

\bibitem{na22}
M. Adamus et al., (EHS/NA22 Collaboration), Z. Phys. C {\bf 37}, 347 (1988)

\bibitem{epzeus}
S. Checkanov et al., (ZEUS Collaboration), Phys. Lett. B {\bf 652}, 1 (2007)

\bibitem{cms}
V. Khachatryan et al., (CMC Collaboration), Phys. Rev. Lett. {\bf 105}, 032001 (2010)

\bibitem{atlas}
G. Aad et al., (ATLAS Collaboration), Eur. Phys. C {\bf 82}, 608 (2022)

\bibitem{alice}
K. Aamodt et al., (ALICE Collaboration), Phys. Rev. D {\bf 84}, 112004 (2011)

\bibitem{jetset}
T.Sj$\ddot{\rm o}$strand, CERN-TH 6488/92, @5035/w5036

\bibitem{hepph9606365}
T. Osada et al. ArXiv:hep-ph/9606365

\bibitem{npb5131998627}
B. Andersson and M. Ringer, Nucl. Phys. B {\bf 513}, 627 (1998)

\bibitem{bolz}
J. Bolz et al., Phys. Rev. D {\bf 47}, 3860 (1993)

\bibitem{plb3511995293}
L. L$\ddot{\rm o}$nnblad, and T. Sj$\ddot{\rm o}$strand, Phys. Lett. B {\bf 351}, 293 (1995)

\bibitem{EurPhysJC21998165}
L. L$\ddot{\rm o}$nnblad, and T. Sj$\ddot{\rm o}$strand, Eur. Phys. J. C {\bf 2}, 165 (1998)

\bibitem{BECinMP}
T.Sj$\ddot{\rm o}$strand, in Multiparticle Production, Eds R. Hwa, (Wold Scientific, Singapore, (1979) p.237

\bibitem{eurphysjc822022608}
G. Aad et al., (ATLAS Collaboration), Eur. Phys. J. C {\bf 82}, 608 (2022)

\bibitem{hybrid}
R. G. Ping et al. Chin. Phys. C {\bf 40}, 113002 (2016).

\bibitem{phokara}
H. Czyz, M. Gunia and J. H. Kuhn, JHEP{\bf 08}, 110 (2013)

\bibitem{conexc}
R. G. Ping et al., Chin. Phys. C {\bf 28}, 083001 (2014)

\bibitem{besiiir}
M. Ablikim et al., (BESIII Collaboration), Phys. Rev. Lett. {\bf 128}, 062004 (2022)

\bibitem{GEANT4}
S. Agostinelli et al., (Geant4 Collaboration), Nucl. Instrum. Meth. A {\bf 506}, 250 (2003)

\end{thebibliography}
\end{document}